%% file: paper.tex
\documentclass{aa}
\usepackage{graphicx,natbib,times,bm,url}
\usepackage{amssymb}
\usepackage{amsmath}
\graphicspath{{./fig/}{./png/}}
\newcommand{\DDt}{{\rm D}_\theta {}}
\input macros

\titlerunning{}
\title{Robustness of oscillatory $\alpha^2$ dynamos in spherical wedges}
\authorrunning{E. Cole et al.}
\author{E.\ Cole\inst{1,2}, A. Brandenburg\inst{2,3,4,5},
P. J.\ K\"apyl\"a\inst{6,1,2} \and M. J.\ K\"apyl\"a\inst{6}
}

\institute{
Department of Physics, Gustaf H\"allstr\"omin katu 2a, PO Box 64,
FI-00014 University of Helsinki, Finland
\and
Nordita, KTH Royal Institute of Technology and Stockholm University,
Roslagstullsbacken 23, 10691 Stockholm, Sweden
\and
Department of Astronomy, AlbaNova University Center,
Stockholm University, 10691 Stockholm, Sweden
\and
JILA and Department of Astrophysical and Planetary Sciences,
Box 440, University of Colorado, Boulder, CO 80303, USA
\and
Laboratory for Atmospheric and Space Physics,
3665 Discovery Drive, Boulder, CO 80303, USA
\and
ReSoLVE Centre of Excellence, Department of Computer Science, Aalto
University,
PO Box 15400, FI-00076 Aalto, Finland
}
\date{\today,~ $ $Revision: 1.106 $ $}
\begin{document}

\abstract{
Large-scale dynamo simulations are sometimes confined to spherical wedge
geometries by imposing artificial boundary conditions at high latitudes.
This may lead to spatio-temporal behaviours that are not representative
of those in full spherical shells.
}{We study the connection between spherical wedge and full spherical shell
geometries using simple mean-field $\alpha^2$ dynamos. 
}
{We solve the equations for a one-dimensional time-dependent mean-field dynamo
to examine the effects of varying the polar angle $\theta_0$ between the
latitudinal boundaries and the poles in spherical coordinates.
We investigate the effects of turbulent magnetic diffusivity
and $\alpha$ effect profiles
as well as different latitudinal boundary conditions to isolate parameter
regimes where oscillatory solutions are found. Finally, we add shear
along with a damping term mimicking radial gradients to study the
resulting dynamo regimes.
}
{
We find that the commonly used perfect conductor boundary condition leads
to oscillatory $\alpha^2$ dynamo solutions
only if the wedge boundary is at least one degree away from the poles.
Other boundary conditions always produce stationary solutions. By
varying the profile of the turbulent magnetic diffusivity alone,
oscillatory solutions are achieved with models extending to the poles,
but the magnetic field is strongly concentrated near the poles and the
oscillation period is very long.
By changing both the turbulent magnetic diffusivity and $\alpha$ profiles
so that both effects are more concentrated toward the equator,
we see oscillatory dynamos with equatorward drift, shorter cycles,
and magnetic fields distributed over a wider range of latitude.
By introducing radial shear and a damping term mimicking radial gradients,
we again see oscillatory dynamos,
and the direction of drift follows the Parker--Yoshimura rule. 
Oscillatory solutions in the weak shear regime are found only in the
wedge case with $\theta_0=1\degr$ and perfect conductor boundaries. 
}
{A reduced $\alpha$ effect near the poles with a turbulent diffusivity
concentrated toward the equator yields oscillatory dynamos with
equatorward migration and reproduces best the solutions in spherical wedges.
For weak shear, oscillatory solutions are obtained only for normal field
conditions and negative shear. Oscillatory solutions become preferred
at sufficiently strong shear.
Recent three-dimensional dynamo simulations producing solar-like
magnetic activity are expected to lie in this range.
}

\maketitle

\section{Introduction}

The Sun's magnetic field is generally believed to be the result of
a turbulent $\alpha\Omega$ dynamo in which differential rotation
plays an important role.
This is referred to as the $\Omega$ effect, and it has long been identified
as a robust mechanism for amplifying the azimuthal magnetic field
of the Sun by winding up the poloidal field \citep{Bab61,UB05,BBBMT10}.
The production of poloidal field, on the other hand, is more complicated
and harder to verify in computer simulations, but it is thought
to be associated with helical motions in the rotating, density
stratified convection zone \citep{Par55,SKR66}. This process is
commonly parametrised by an $\alpha$ effect.
Although there remain substantial uncertainties regarding the $\alpha$
effect as an important ingredient at large magnetic Reynolds numbers
\citep{CH06}, simulations of turbulence
and rotating convection have subsequently confirmed that conventional
estimates of $\alpha$ and turbulent diffusivity $\etat$ are reasonably
accurate up to moderate values of the magnetic Reynolds number
\citep{SBS08,KKB09}.

Simulations also demonstrate the generation of differential rotation
from anisotropic rotating convection, which amounts to a relative value
of 20--30\% in latitude \citep[e.g.][]{Mie00,KMGBC12}.
However, whether or not this is enough to drive an $\alpha\Omega$ dynamo
as opposed to an $\alpha^2$ dynamo, in which the $\Omega$ effect would be
subdominant, can only be decided on the basis of quantitative calculations.

In the absence of a conclusive answer, one tends to resort to qualitative
arguments.
One is related to the clear east--west orientation of bipolar regions in
the Sun, which suggests that the azimuthal field must be much stronger
than the poloidal field.
Another argument is that $\alpha\Omega$ dynamos are usually cyclic
and can display equatorward migration of magnetic field either through
suitable radial differential rotation \citep{Par55,SK69} or through
sufficiently strong meridional circulation in the presence of an
$\alpha$ effect that operates only in the surface layers \citep{CSD95}.
However, both arguments are problematic.
Although it is probably true that the azimuthal field is stronger than
the poloidal, their ratio may not be large enough to justify the dominance
of the $\Omega$ effect.
Furthermore, $\alpha^2$ dynamos may well be oscillatory
\citep[e.g.][]{KMB13,MS14} and can display
equatorward migration under suitable conditions \citep{MTKB10}.
A completely different argument that motivates the study of oscillatory
$\alpha^2$ dynamos are recent simulations of convective dynamos in
spherical wedges and full shells that also show equatorward migration
\citep{KMB12,KMCWB13,WKMB13,ABMT13}.
It is now believed that the equatorward migration in the
simulations is facilitated by a region of negative shear and positive
(negative) $\alpha$ effect in the northern (southern) hemisphere -- in
accordance with the Parker--Yoshimura rule \citep{WKKB14}.
Recently an alternative scenario was reported by \cite{DWBG15}, who
found that the sign of the $\alpha$ effect can be inverted in certain
parameter ranges allowing equatorward migration also with positive
radial shear.
Although it is unclear to what extent those simulations represent stellar
magnetic fields, it might be helpful to understand first the mechanism
operating in those simulations before trying to understand real stars.

While the idea of explaining equatorward migration through $\alpha^2$
dynamo action might work in spherical wedge simulations, there is the
problem that such solutions have never been seen in full shell simulations
that extend not just to high latitudes, but go all the way to the poles.
Indeed, $\alpha^2$ dynamos in full spherical shells are known
to be steady \citep{SK69b}.
Exceptions are dynamos with an anisotropic $\alpha$ tensor
\citep{REO03} and the non-axisymmetric oscillatory solutions found by
\cite{JW06}, but for an isotropic $\alpha$ effect, oscillatory
axisymmetric $\alpha^2$ dynamos seem to be an artefact of having
imposed a boundary condition at high latitudes.
One could choose another boundary condition; a normal-field
(pseudo-vacuum) boundary condition might be an obvious choice, but from
corresponding Cartesian simulations we know that this would again lead
to oscillatory solutions, but with poleward migration \citep{BCC09}.

Although the mean-field description of oscillatory $\alpha^2$ dynamos
seems to face an internal inconsistency regarding the limit to full spherical
shells, there remains the question whether certain changes in the setup
of the full spherical shell model could lead to oscillatory solutions
that are internally consistent and otherwise similar to the solutions
in spherical wedges.
There is a priori no physical motivation for this, but from a mathematical
point of view, this is a natural choice when trying to reproduce the
conditions encountered previously with a perfect conductor boundary condition.
One possibility is a suitable latitudinal $\etat$ profile with a
larger conductivity (weaker magnetic diffusion) at high latitudes to
simulate the behaviour of perfect conductor boundary conditions used in
spherical wedges.

In each of those cases, it is important to assess how much shear would be
needed to change the dynamo mode into an $\alpha\Omega$ type mode.
To keep things simple, we employ a one-dimensional model with
only latitudinal extent.
However, in its standard formulation, with radial derivatives simply being
dropped, the first excited mode of such an $\alpha\Omega$ dynamo is
non-oscillatory \citep{JBMT90}.
This is an artefact that is easily removed by substituting radial
derivatives by a damping term \citep{KS95,MSKP04}, instead of setting
them to zero.

We begin by describing our model in detail, next focus on the
analysis of spherical wedges of different extent and turn then to
full spherical shells with variable latitudinal $\etat$ profiles.
In view of the aforementioned complications
regarding the possibility of oscillatory behaviour in the corresponding
$\alpha\Omega$ dynamos, we also discuss the sensitivity of our
solutions with respect to an additional damping term that mimics the
otherwise neglected radial derivative terms.

\section{Model}

We consider the mean-field dynamo equation for the mean magnetic field
$\meanBB$ with a given mean electromotive force $\meanEMF$ in the form
\EQ
{\partial\meanBB\over\partial t}=\nab\times\left(
\meanUU\times\meanBB+\meanEMF-\eta\mu_0\meanJJ\right),
\label{dmeanBBdt}
\EN
where $\meanUU=\pphi\varpi\Omega$ is the mean flow from angular velocity
with $\varpi=r\sin\theta$ being the distance from the axis,
$\Omega(r,\theta)$ is the internal angular velocity,
$\pphi$ is the unit vector in the azimuthal direction,
$\meanJJ=\nab\times\meanBB/\mu_0$ is the mean current density,
$\mu_0$ is the vacuum permeability, and $\eta$ is the non-turbulent
magnetic diffusion coefficient.
In the absence of a memory effect, and under the assumption of isotropic
$\alpha$ effect and turbulent magnetic diffusivity $\etat$, the mean
electromotive force is given by
\EQ
\meanEMF=\alpha\meanBB-\etat\mu_0\meanJJ.
\label{meanEMF}
\EN

We solve \Eqs{dmeanBBdt}{meanEMF} numerically using sixth-order finite
differences in space and a third-order accurate time-stepping scheme.
We employ the {\sc Pencil Code}\footnote{http://pencil-code.github.com/},
which solves the governing equations in terms of the mean magnetic
vector potential $\meanAA$, such that $\meanBB=\nab\times\meanAA$.
It is convenient to use the advective gauge \citep{BNST95,CHBM11},
in which the electrostatic
potential has a contribution $\meanU_\phi\meanA_\phi$, so that
\EQ
{\partial\meanAA\over\partial t}=
-\varpi\meanA_\phi\nab\Omega+\meanEMF-\eta\mu_0\meanJJ.
\label{dmeanA1}
\EN
To allow for the use of a one-dimensional model with
$\meanBB=\meanBB(\theta,t)$, we restrict ourselves
to an angular velocity profile that varies linearly in $r$,
i.e., $\Omega(r,\theta)=rS(\theta)$, so the angular velocity gradient
becomes $\nab\Omega=(S,\partial_\theta S,0)$.
The mean current density is then
\EQ
\meanJJ=\mu_0^{-1}R^{-2}\left(
\DDt \meanA_\theta -\DDt\partial_\theta \meanA_r,\;
\partial_\theta \meanA_r,\;
-\partial_\theta \DDt \meanA_\phi\right),
\EN
where ${\rm D}_\theta =\cot\theta+\partial_\theta$ is a modified
$\theta$ derivative.
To account for the neglect of $r$ derivatives, we add in \Eq{dmeanA1}
a damping term of the form $-\mu^2\meanAA$, i.e., we have
\EQ
{\partial\meanAA\over\partial t}=
-\varpi\meanA_\phi\nab\Omega+\meanEMF-\eta\mu_0\meanJJ-\mu^2\meanAA
\quad\mbox{(with $\partial_r=0$)};
\EN
see \cite{MSKP04} for a survey of solutions for different values of $\mu$.
For $\alpha$ and $\etat$ we use latitudinal profile functions of the form
\EQ
\alpha=\alpha_0\cos\theta\left(a_0+a_2\sin^2\!\theta+\ldots+a_n\sin^n\!\theta\right),
\EN
\EQ
\etat=\etatz\left(e_0+e_2\sin^2\!\theta+\ldots+e_n\sin^n\!\theta\right),
\label{CondProf}
\EN
where $a_i$ and $e_i$ are coefficients denoted by the vectors
$\bm{{a}}=(a_0,a_2,a_4,\ldots,a_n)$ and
$\bm{{e}}=(e_0,e_2,e_4,\ldots,e_n)$, respectively. However, we
often refer to only the three first components as
$\aaaa=(a_0,a_2,a_4)$ and $\ee=(e_0,e_2,e_4)$.
These expansions can also be expressed in terms of Legendre polynomials,
which are orthonormal functions that obey regularity at the poles.
The occurrence of higher order terms in $\alpha$ has been associated with
higher orders terms in $\grav\cdot\OO$, which are normally omitted in
theoretical calculations \citep{RB95}.

As usual, the problem is governed by two dynamo numbers,
\EQ
C_\alpha=\alpha_0 R/\etatz,\quad
C_\Omega=S_0 R^2/\etatz,
\EN
where $S(\theta)=S_0$ is now a constant.
We consider the following sets of boundary conditions:
\EQ
\partial_\theta\meanA_r=\meanA_\theta=\meanA_\phi=0
\quad\mbox{(SAA; regularity on $\theta=0$)},\quad
\label{REG}
\EN
\EQ
\meanA_r=\partial_\theta \meanA_\theta=\meanA_\phi=0
\quad\mbox{(ASA; perf.\ cond.\ on $\theta=\theta_0$)},\quad
\label{PC}
\EN
\EQ
\partial_\theta \meanA_r=\meanA_\theta=\partial_\theta\meanA_\phi=0
\;\;\mbox{(SAS; normal field on $\theta=\theta_0$)},\;
\label{NF}
\EN
where the sequence of letters S and A refer respectively to symmetric
($\partial_\theta=0$) and antisymmetric (vanishing function value)
of $\meanA_r$, $\meanA_\theta$, and $\meanA_\phi$ across the boundary.
The same conditions are also applied on the corresponding boundary
in the southern hemisphere where $\pi-\theta=\theta_0$.
In this work, no symmetry condition on the equator is applied,
so the parity of the solution is not constrained.

As initial conditions, we assume a seed magnetic field
consisting of low-amplitude Gaussian noise.
Such a field is sufficiently complex so that the fastest growing
eigenmode of either parity tends to emerge after a short time.
Note that mixed parity solutions are only possible in the nonlinear
regime \citep{BKMMT89}, but this will not be considered here.

\section{Results}

We consider separately the cases where the dynamo is driven either
solely by the $\alpha$ effect ($\alpha^2$ dynamos) or by the
combined action of $\alpha$ effect and large-scale shear
($\alpha^2\Omega$ dynamos).

\subsection{$\alpha^2$ dynamos}

\subsubsection{Varying $\theta_0$}
\label{SecVaryTheta}

We begin by considering the simplest case with $\aaaa=(1,0,0)$ and
$\ee=(1,0,0)$ resulting a spatially constant turbulent diffusivity and
a $\cos\theta$ profile for $\alpha$
We have calculated the critical value of $C_\alpha$,
hereafter $\Calp$, for an oscillatory
$\alpha^2$ dynamo, i.e., where $C_\Omega=0$.
We used the boundary conditions SAA (Eq.~\ref{REG}),
ASA (Eq.~\ref{PC}),
and SAS (Eq.~\ref{NF}) for selected values of $\theta_0$.

\begin{figure}[t!]\begin{center}
\includegraphics[width=\columnwidth]{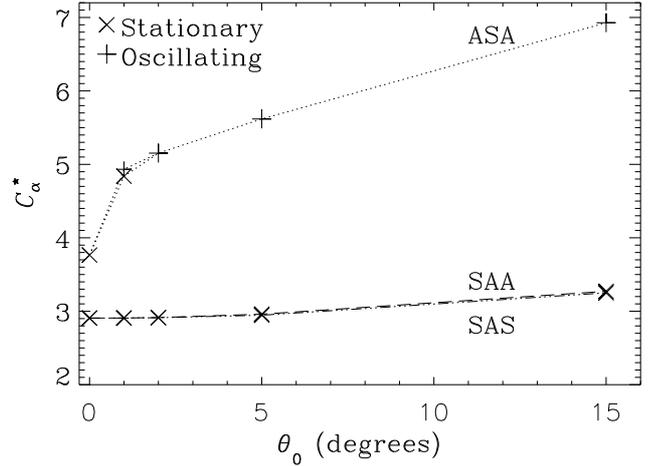}
\end{center}\caption[]{
Dependence of $\Calp$ on $\theta_0$
for the three boundary conditions ASA, SAA, and SAS.}
\label{theta0_alpcrit}
\end{figure}

\begin{figure}[t!]\begin{center}
\includegraphics[width=\columnwidth]{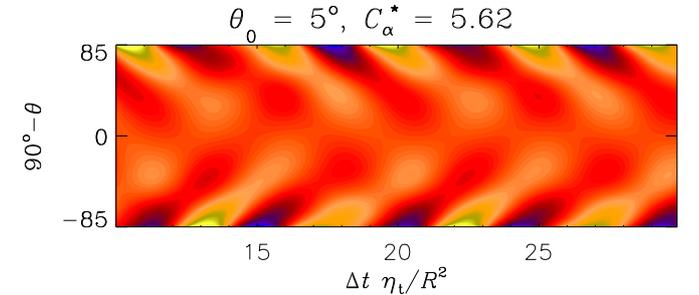}
\end{center}\caption[]{
Azimuthal magnetic field $\mean{B}_\phi$ for an oscillatory dynamo
with $\theta_0=5\degr$ and the boundary condition ASA.}
\label{bfly_5deg}
\end{figure}

It turns out that $\Calp$ decreases as we
approach the pole ($\theta_0\to0$); see \Fig{theta0_alpcrit}.
The SAA and SAS
boundary conditions result in very similar non-oscillatory solutions
with a $\Calp$ of only approximately 40\% of that $\Calp$ 
obtained for the ASA boundary condition. 
Oscillatory solutions show travelling waves that propagate equatorward;
see \Fig{bfly_5deg}.
The boundary
condition with the greatest variation of $\Calp$ with $\theta_0$
is the perfect conductor, ASA. 
We also find that the most easily excited dynamo mode changes from
stationary to oscillatory as $\theta_0$ increases from zero to one
degree in that case.
For the case where $\theta_0=1\degr$ we find both stationary and
oscillatory solutions depending on the initial conditions. The
critical dynamo number is slightly higher for the oscillatory mode
than the corresponding value of the stationary solution.

These results suggest that we cannot regard the limit $\theta_0\to0$
with isotropic $\alpha$ effect and turbulent diffusion and perfect
conductor boundaries as an approximation to a full spherical shell model
when searching for oscillatory solutions.
Extending the model to the poles with the ASA boundary condition 
changes the resulting dynamo from oscillatory to stationary. 
The SAA and SAS boundary conditions give stationary solutions with 
relatively similar values for $\Calp$, 
but the ASA boundary condition 
near the poles gives both oscillatory and stationary solutions, 
depending on the initial conditions of the seed magnetic field.
While no stationary solutions were found for $\theta_0 >1\degr$,
their existence is not ruled out by our models.

\subsubsection{Varying latitudinal $\etat$ profile}

Given that we have found the limit $\theta_0\to0\degr$ 
in the case of the perfect conductor boundary condition 
not to be an
approximation to a full spherical shell model, we now investigate
whether physically motivated alterations of the full spherical
shell model with the SAA boundary condition could produce 
oscillatory, equatorward solutions similar to those found 
for $\theta_0\neq0\degr$ with the ASA boundary condition.
An obvious possibility is the use of an $\etat$ profile that corresponds
to high conductivity near the pole.
Such a profile could correspond to the possible 
effect of rotation on the magnetic diffusivity \citep{KRP94}
at various latitudes. 

\begin{table}
\caption{ 
$\Calp$ for pure $\alpha^2$ dynamos with varied magnetic diffusivity and
$\alpha$ profiles and the corresponding oscillations frequencies
in units of $\etatz/R^2$.
}\begin{tabular}{ll|lll}
\hline \hline
  &  & & $\aaaa$  & \\ 
&  & $(1,0,0)$ & $(0,1,0)$ & $(0,0,1)$ \\
$e_i$ & $e_0$ &  \ $\Calp \ \vert\  
\omega$ & \ $\Calp \ \vert\ \omega$
 & \ $\Calp  \ \vert\ \omega$ \\
\hline \hline
$e_2$ & 0.01 & 0.236 $\vert$ ----  & 4.063 $\vert$ 0.405 & 9.532 
$\vert$  0.562\\
$e_2$ & 0.05 & 0.558 $\vert$ ----  & 5.308 $\vert$ 0.288 & 11.39 
$\vert$ 0.654\\
\hline
$e_4$ & 0.01 & 0.096 $\vert$ 0.008 & 1.045 $\vert$ 0.207 & 4.144 
$\vert$ 0.298\\
$e_4$ & 0.05 & 0.326 $\vert$ ---- & 2.587 $\vert$ 0.332 & 7.039 
$\vert$ 0.548\\
\hline
$e_6$ & 0.01 & 0.070 $\vert$ 0.005 & 0.541 $\vert$ 0.184 & 2.175  
$\vert$ 0.215\\
$e_6$ & 0.05 & 0.265 $\vert$ ---- & 1.733 $\vert$ 0.258 & 4.857 
$\vert$ 0.419\\
\hline
$e_8$ & 0.01 & 0.059 $\vert$ 0.003 & 0.403 $\vert$ 0.131 & 1.463 
$\vert$ 0.165\\
$e_8$ & 0.05 & 0.238 $\vert$ ---- & 1.384 $\vert$ 0.199 & 3.727 
$\vert$ 0.364\\
\hline
\end{tabular}
\label{vary_cond_prof}
\end{table}

One possible alteration to the diffusivity profile is 
to use higher order terms for $\etat$.
In particular, we examine solutions where the orders $i = 2$,
4, 6, and 8 are used for $e_i$; see Eq.~(\ref{CondProf}).
Solutions are examined for $e_0 = \eta / \etatz = 0.01$ and $0.05$.
A non-zero uniform value of $\eta$ is needed to ensure the
stability of the solutions in the cases where the turbulent
magnetic diffusivity is zero at the poles due to the profiles being
proportional to powers of $\sin\theta$, which vanishes at the poles.

Neither value of $\eta$ used here leads to spurious growth in the
absence of an $\alpha$-effect.
Furthermore, we calculate the oscillation frequency as $\omega =
2\pi/T$ where $T$ is the period of oscillation for the large-scale
magnetic field.

\begin{figure}[t!]\begin{center}
\includegraphics[width=\columnwidth]{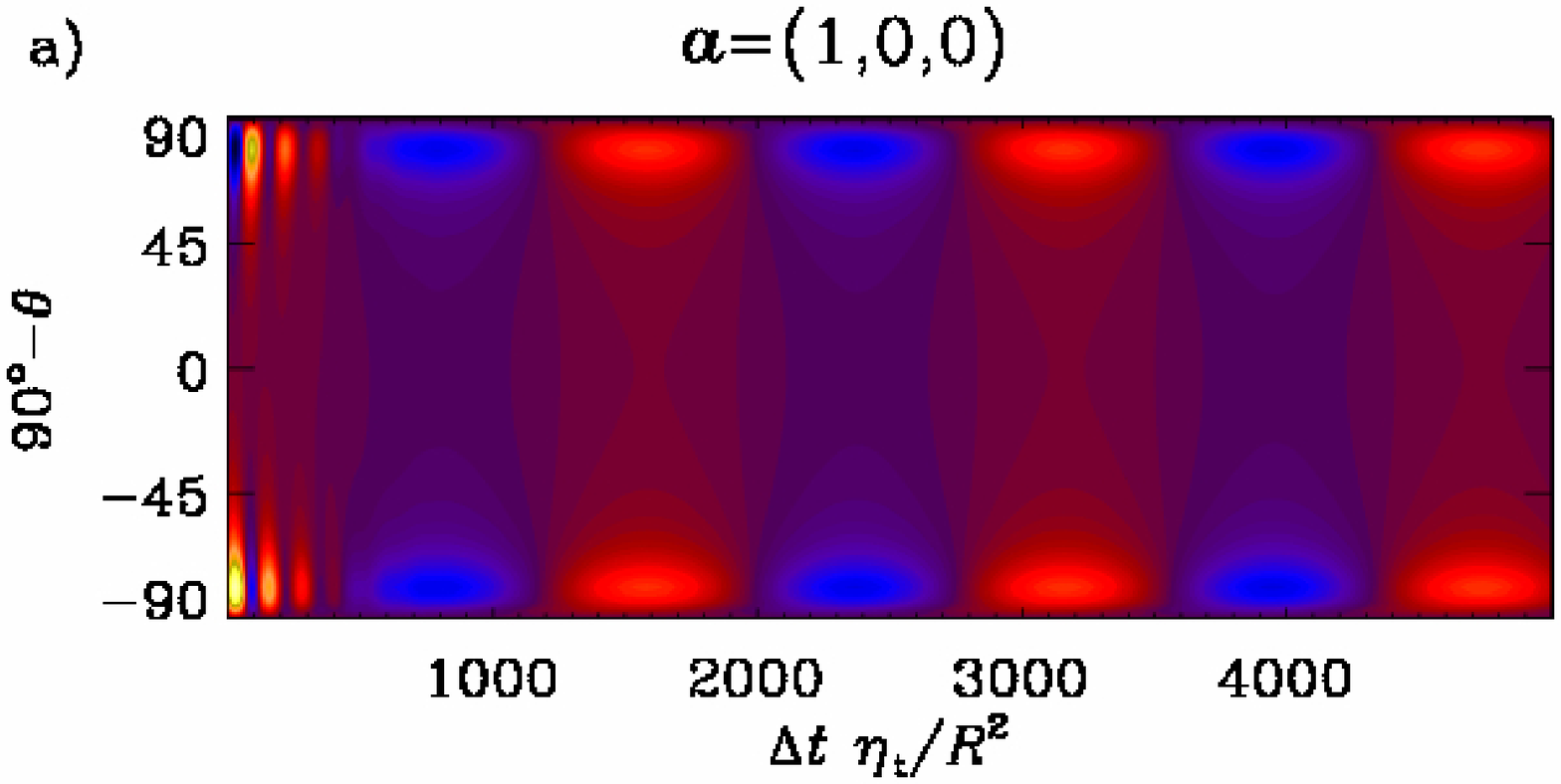}
\includegraphics[width=\columnwidth]{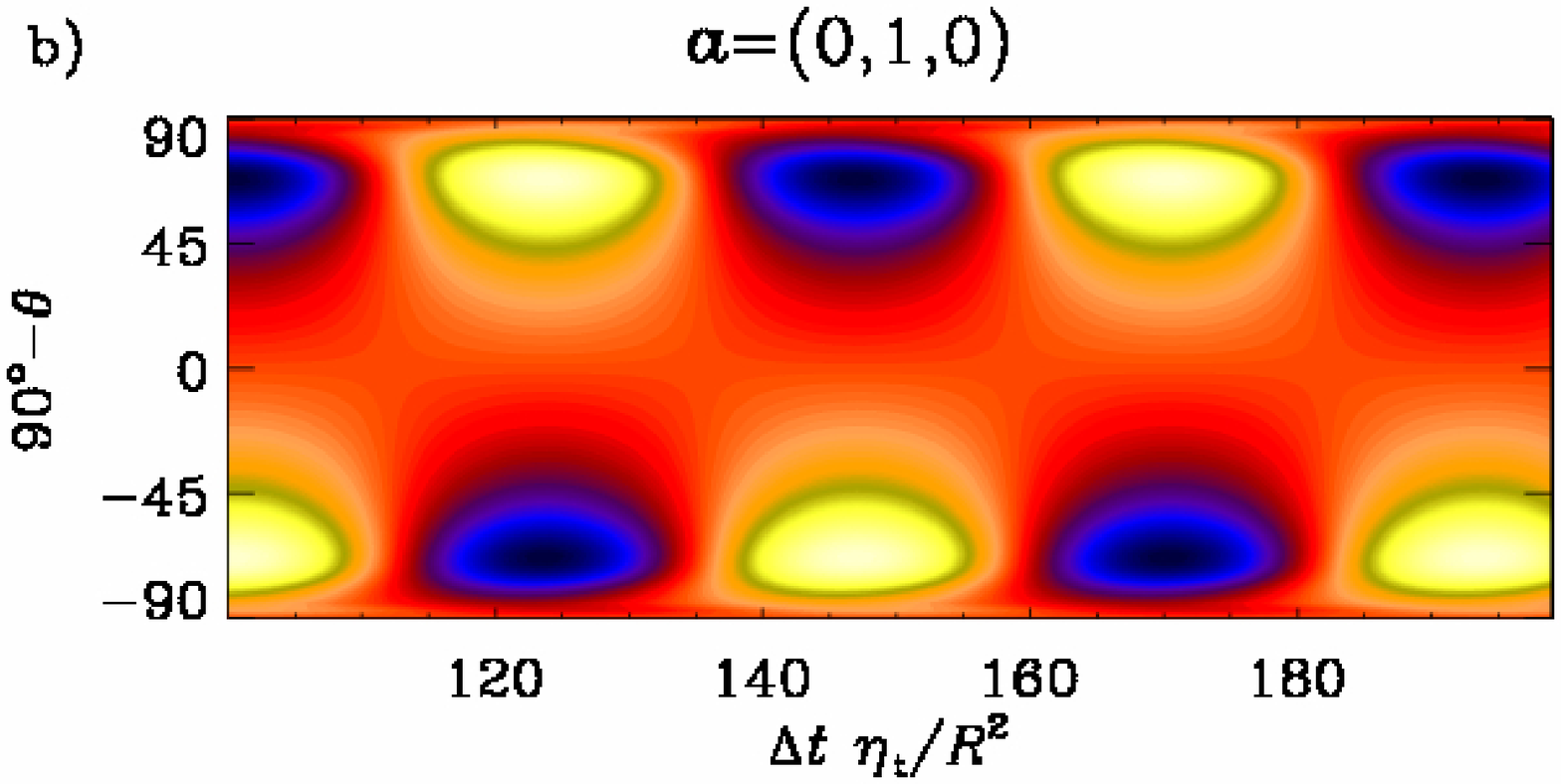}
\includegraphics[width=\columnwidth]{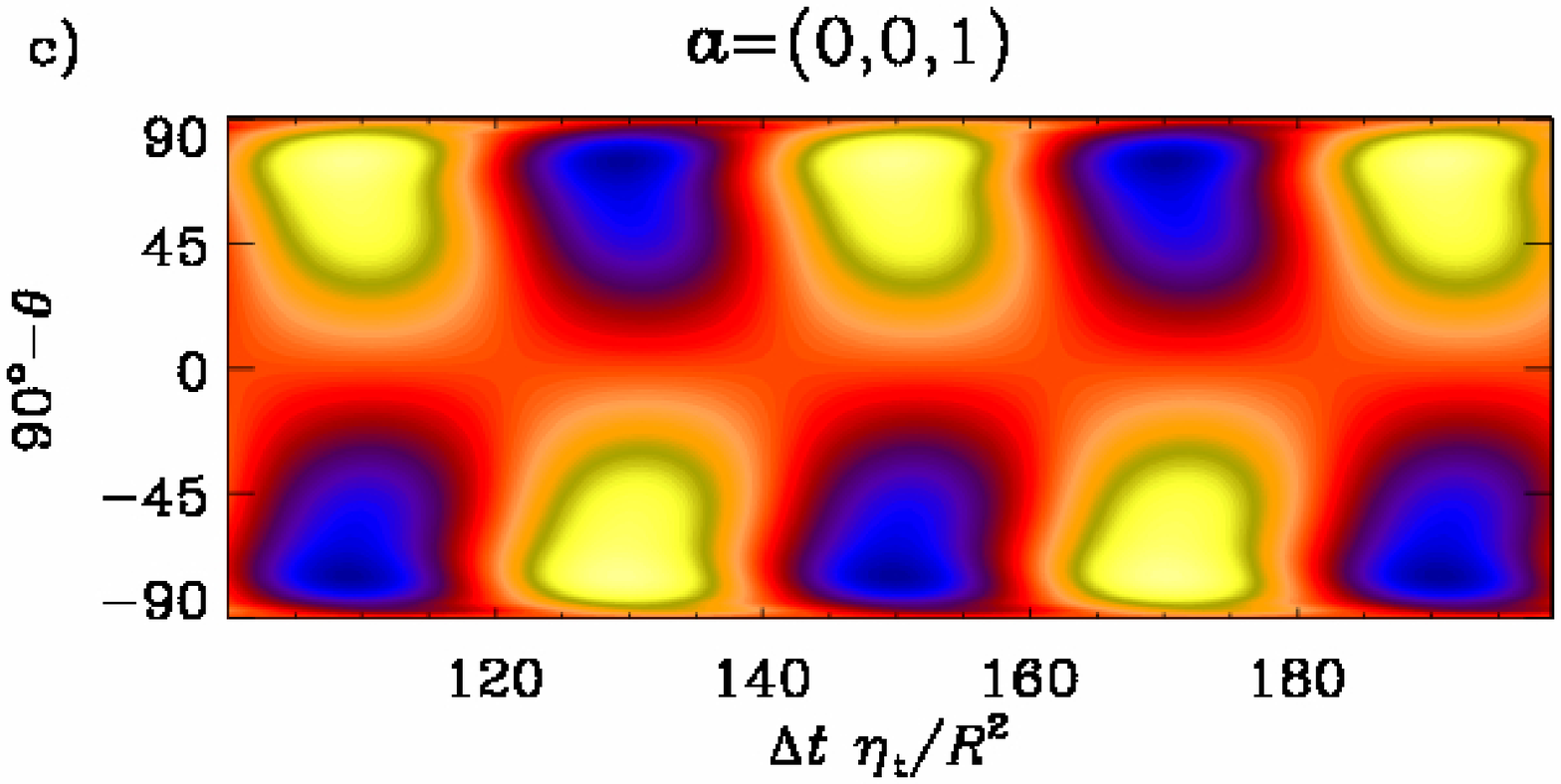}
\end{center}\caption[]{
Azimuthal magnetic field for $e_4$, $e_0=0.01$
in \Tab{vary_cond_prof} with $\theta_0=0\degr$ 
and the SAA condition. 
}\label{bfly_0deg_sin4y_3}\end{figure}

Values for $\Calp$ are indicated in Table \ref{vary_cond_prof} for
cases where the turbulent diffusivity and $\alpha$ effect profiles
are expanded up to orders $e_8$ and $a_4$, respectively.
We find that for $\aaaa = (1,0,0)$, the $e_0 = 0.05$ case
produces only stationary solutions, but at $e_0 = 0.01$,
only solutions for $n = 2$ are stationary and all higher orders oscillate;
see \Tab{vary_cond_prof}.
Some solutions initially show rapidly oscillating behaviour,
exhibiting antisymmetry with respect to the equator, but these
disappear later and only a slower, persistent oscillatory mode
remains: see the top panel of \Fig{bfly_0deg_sin4y_3}.
These low-frequency oscillations have neither equatorward nor 
poleward migration, and are symmetric about the equator. 
$\Calp$ increases with $e_0$,
and decreases as $n$ increases for $e_n$, in accordance with the total
diffusion increasing and decreasing, respectively.
The frequency of the oscillatory modes
found for  $e_0 = 0.01$ decreases as $n$ increases.
This is also consistent with mean-field theory where the oscillation
frequency is proportional to the magnetic diffusion coefficient.
The magnetic field is antisymmetric with respect to the equator in all
cases, except for $\aaaa = (1,0,0)$ and $e_0=0.01$; see the
top panel of \Fig{bfly_0deg_sin4y_3}

\begin{figure}[t!]\begin{center}
\includegraphics[width=\columnwidth]{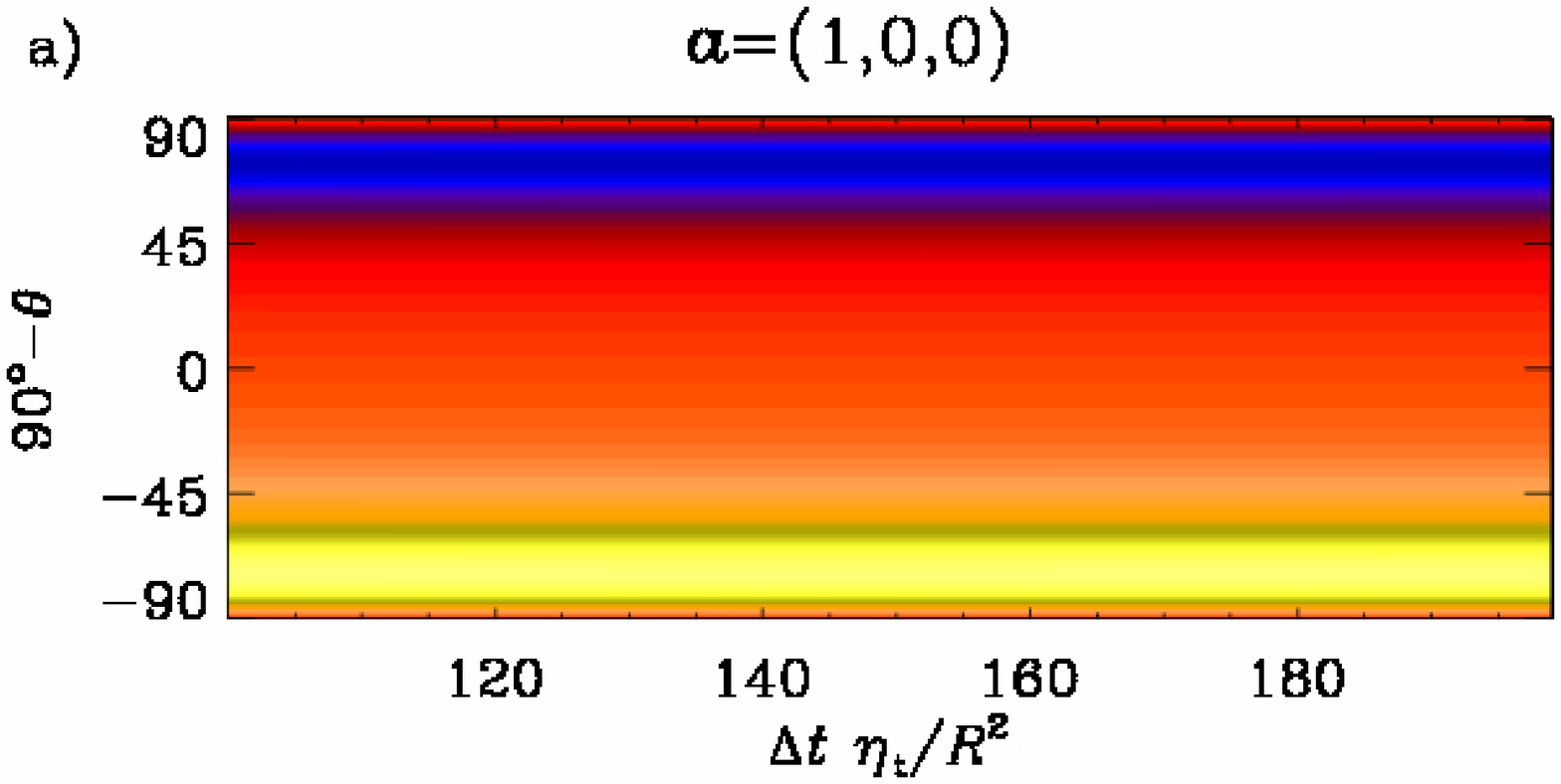}
\includegraphics[width=\columnwidth]{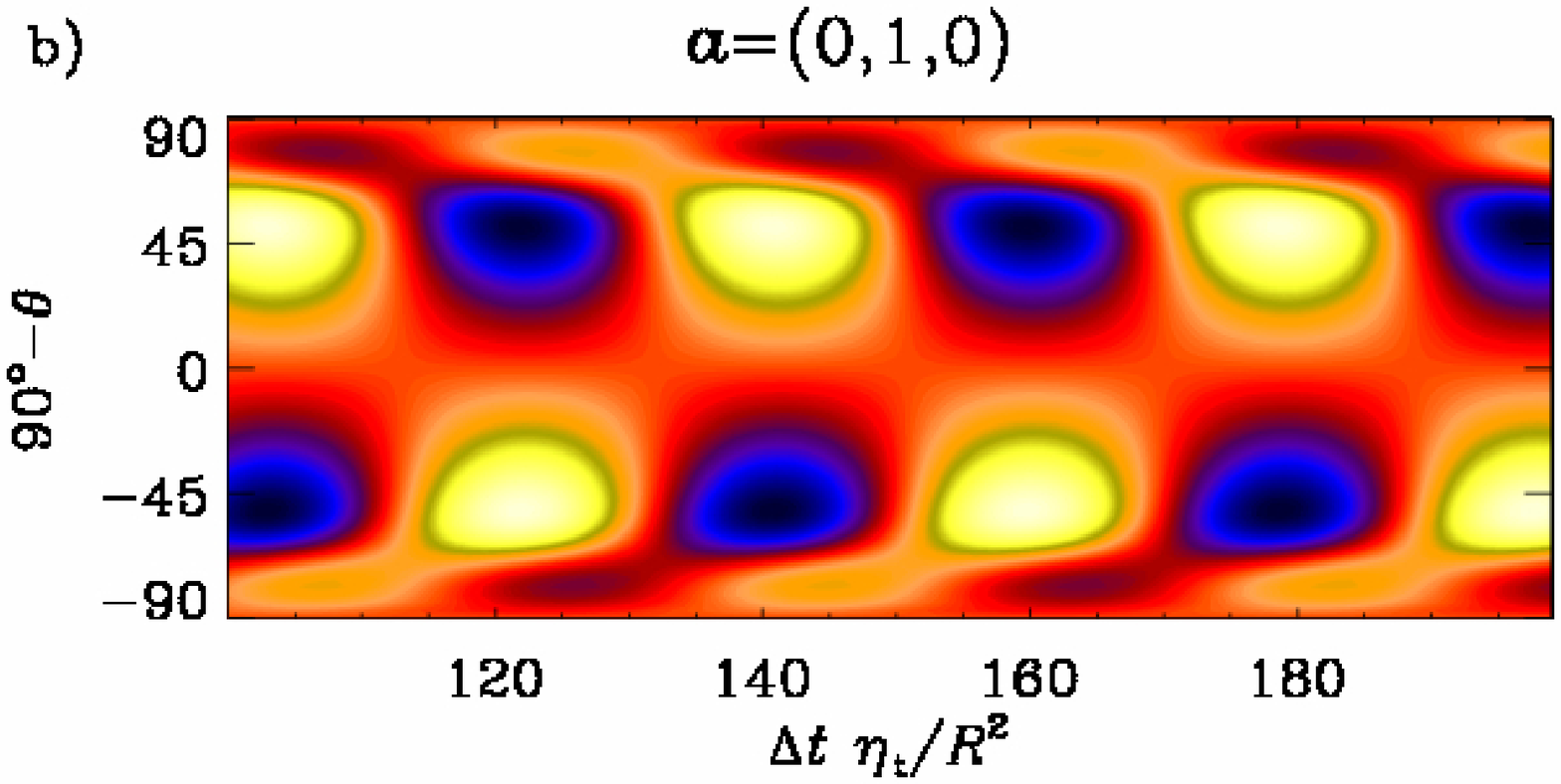}
\includegraphics[width=\columnwidth]{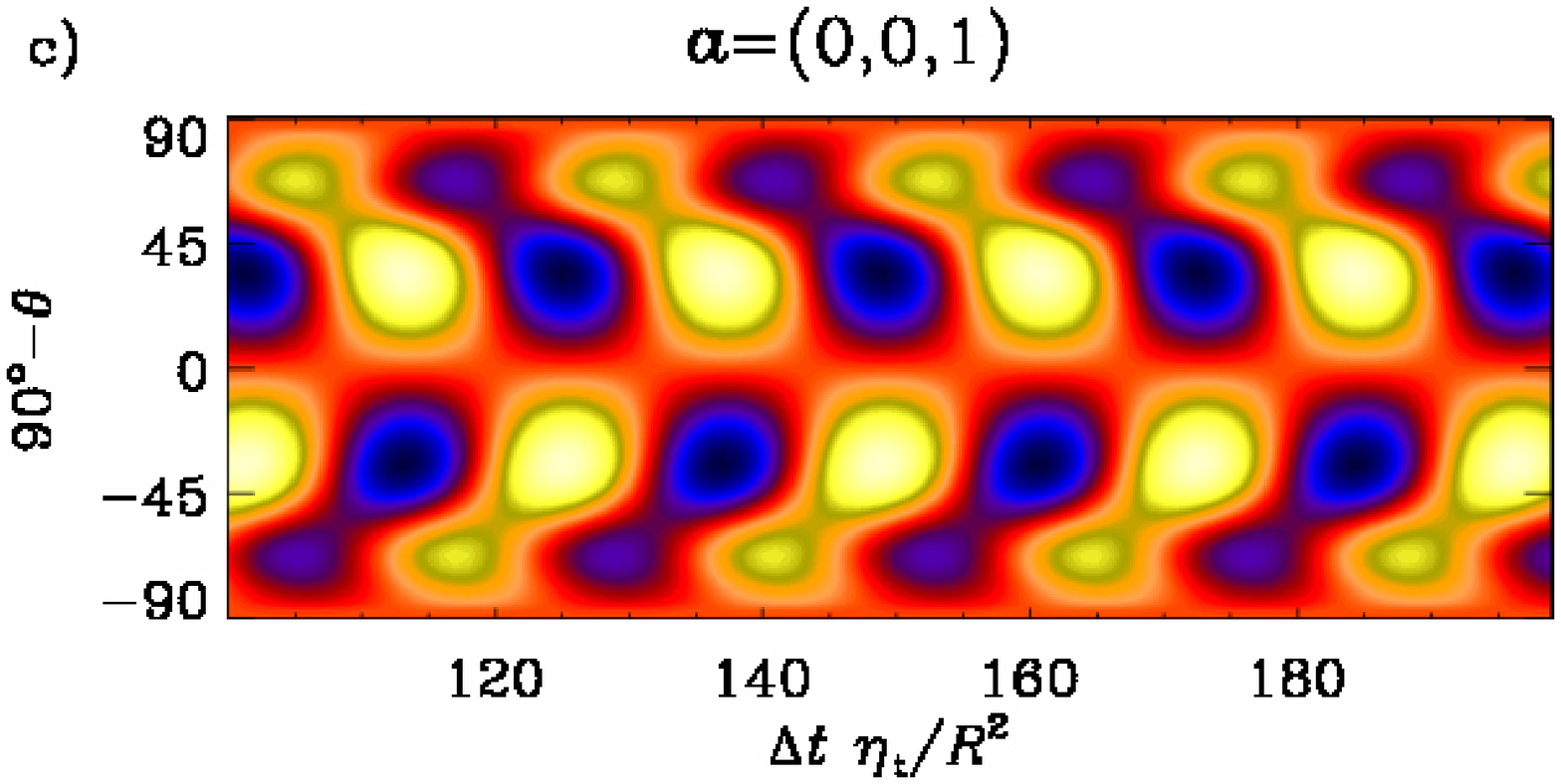}
\end{center}\caption[]{
Azimuthal magnetic field for $e_4$, $e_0=0.05$
in \Tab{vary_cond_prof} with $\theta_0=0\degr$ 
and the SAA condition.
}\label{bfly_0deg_sin4y_1}\end{figure}

The azimuthal magnetic field is strongly concentrated toward the poles
when the $\alpha$ effect has only the $\cos \theta$ variation in
latitude; see the top panels of \Figs{bfly_0deg_sin4y_3}{bfly_0deg_sin4y_1}.
In view of the equatorial magnetic field concentration in the Sun and in
three-dimensional solar dynamo simulations, where the kinetic helicity
is known to be strongly concentrated toward the equator \citep{KMB12},
it is of interest
to consider models with $\aaaa=(0,1,0)$ and $\aaaa=(0,0,1)$, so that
the $\alpha$ effect is more concentrated toward lower latitudes. 
Indications for $\alpha$ being stronger at lower 
latitudes have been observed, for example, 
in models of rapidly rotating convection \citep{KKOS06}. 
The values for $\Calp$ are given in \Tab{vary_cond_prof},
columns for $\aaaa=(0,1,0)$ and $\aaaa=(0,0,1)$.
A similar trend as for the case where $\aaaa = (1,0,0)$ is seen, 
where higher orders of $e_i$ result in lower values for $\Calp$,
in accordance with lower total diffusion.
Changes in the $\alpha$ profile have a larger effect on 
$\Calp$ than changes in the diffusivity profile. 
However, this is simply because, owing to the presence of the
$\cos\theta$ factor in the $\alpha$ profile, its maximum value
diminishes as higher powers of $\sin \theta$ are used, while the
maximum value of $\etat$ is always unity, irrespective of the profile.
The oscillation frequencies of the solutions for 
$\aaaa=(0,1,0)$ and $\aaaa=(0,0,1)$ are two orders of magnitude 
higher than the low-frequency mode seen for $\aaaa=(1,0,0)$.
It turns out that the magnetic field is then more uniformly distributed
over all latitudes; see \Figs{bfly_0deg_sin4y_3}{bfly_0deg_sin4y_1}.
For $e_0 = 0.01$, this distribution is largely uniform
with very slight equatorward drift (\Fig{bfly_0deg_sin4y_3},
middle and bottom), and when $e_0= 0.05$,
the equatorward drift becomes more pronounced and 
extends to lower latitudes (middle and bottom panels of
\Fig{bfly_0deg_sin4y_1}).

In summary, extending the model all the way to the poles and including
an $\etat$
profile concentrated toward the equator results in oscillatory
behaviour with long cycles but no equatorward migration.
Including an $\alpha$-effect also concentrated at lower latitudes
produces equatorward cycles with shorter cycle periods with strongest
magnetic fields appearing at lower latitudes. These results are in
qualitative agreement with direct and large-eddy simulations
\citep{KMB12,KMCWB13,ABMT13,DWBG15}.

\subsection{$\alpha^2\Omega$ dynamos}

\begin{figure}[t!]\begin{center}
\includegraphics[width=\columnwidth]{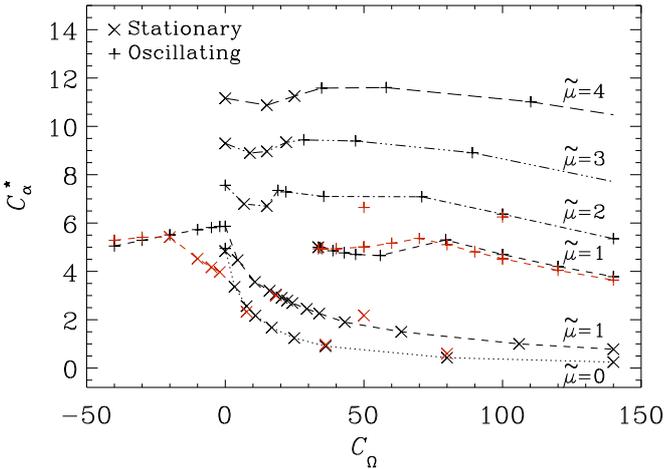}
\end{center}\caption[]{
Values for $\Calp$ as a function of 
$C_\Omega$ for oscillatory (pluses) and stationary (crosses) solutions
for $\theta_0=1\degr$ (black) and $\theta_0=0\degr$ (red).
  }
\label{shear_alpcrit}
\end{figure}

\begin{figure}[t!]\begin{center}
\includegraphics[width=\columnwidth]{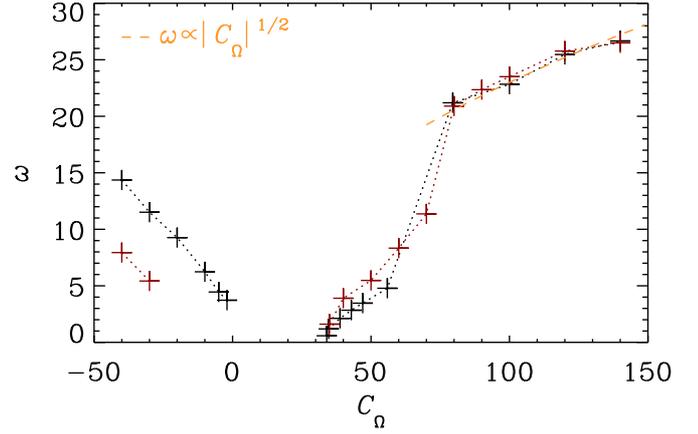}
\end{center}\caption[]{
Angular frequency $\omega$ as a 
function of $C_\Omega$ for $\theta_0=1\degr$ (black) and
$\theta_0=0\degr$ (red) for $\tilde{\mu}=1$.}
\label{com_freq}
\end{figure}

\subsubsection{Overall behaviour of dynamo solutions}
\label{sec:overall}

We now add large-scale radial shear and a damping term given by
$\mu R^2/\etat$
and use $\tilde{\mu}$ to denote $\mu R^2/\etatz$. 
We first explore the dynamo regimes and the dependency on 
$\tilde{\mu}$ by setting $\theta_0 = 1\degr$ and once again use $\aaaa =
(1,0,0)$ and $\ee=(1,0,0)$.
The critical value $\Calp$ now depends on the value of 
$\tilde{\mu}$;
see Fig.~\ref{shear_alpcrit}, black symbols. 
We now concentrate on studying the dynamo modes that are excited in
the system
for values of $\tilde{\mu}$ between 0 and 4
and various values of $C_\Omega$.

When $\tilde{\mu}=0$, all resulting dynamos are stationary, 
with the exception of the case where $C_\Omega = 0$
where oscillations depend on initial conditions,
and $\Calp$ decreases as $C_\Omega$ increases. 
For solutions pertaining to $\tilde{\mu} = 1$, 
two solutions exist in the regime $C_\Omega
\ga 33.5$ with either oscillatory or stationary magnetic
fields.
When $C_\Omega$ is less than this value, 
we find only stationary solutions.
Near this limit, the frequency of oscillations is sensitive to both
$\Calp$ and $C_\Omega$ and even small changes can double the frequency. 
The $\Calp$ for stationary dynamos is significantly 
less than for oscillating solutions.
It is possible that for $\tilde{\mu} > 1$ a similar bifurcation also
exists, as there always appears a jump in $\Calp$ as the dynamo mode
changes from stationary to oscillatory. However, at least in the case
with $\tilde{\mu}=2$, the stationary solutions were found to
disappear.
For cases where $\tilde{\mu} > 2$, $\Calp$ decreases with 
$C_\Omega$, and oscillations only occur above  
certain critical values for $C_\Omega$.
In the regime of negative shear ($C_\Omega < 0$), 
all solutions found were oscillatory.
 
\begin{figure}[t!]\begin{center}
\includegraphics[width=\columnwidth]{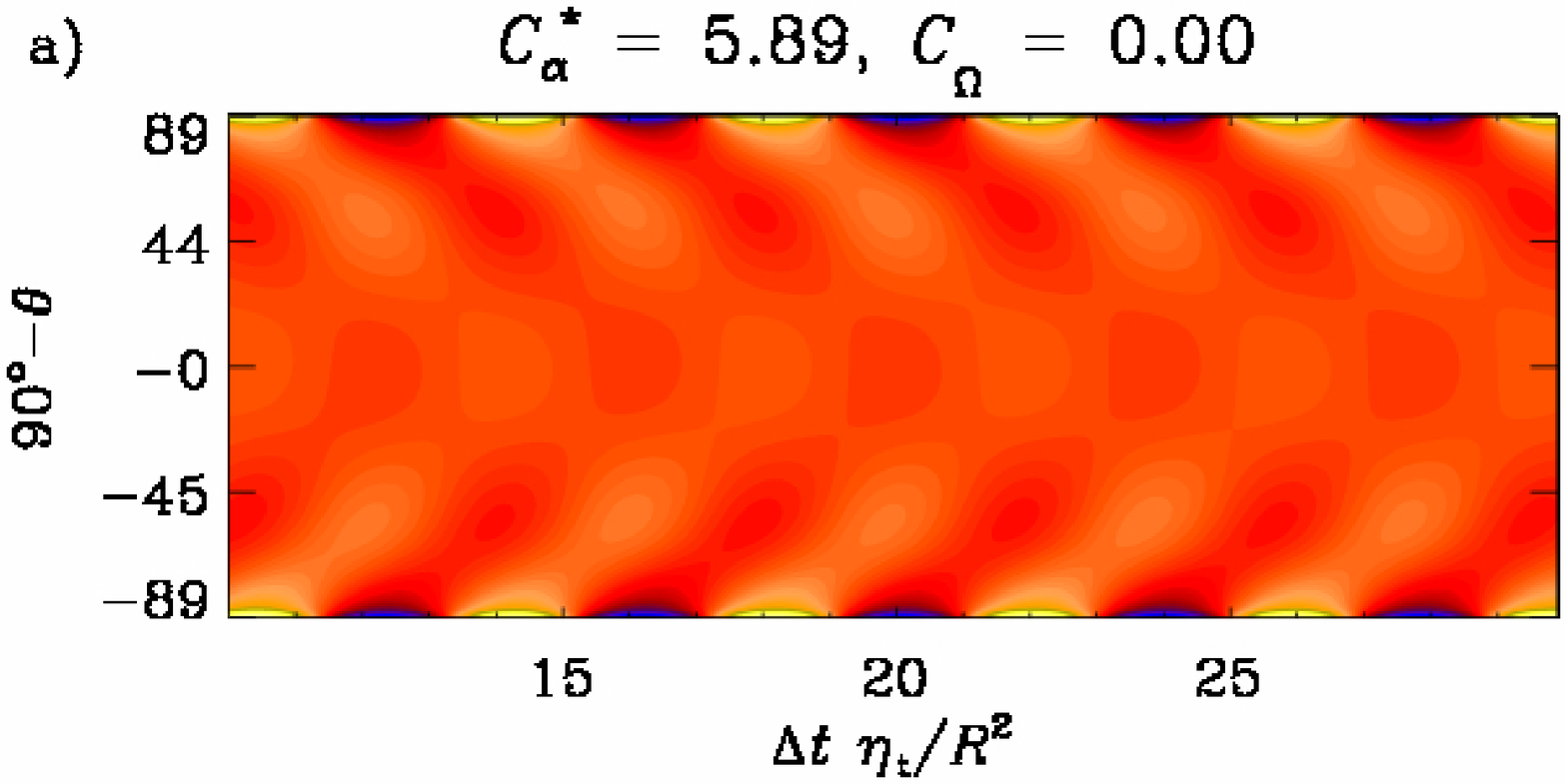}
\includegraphics[width=\columnwidth]{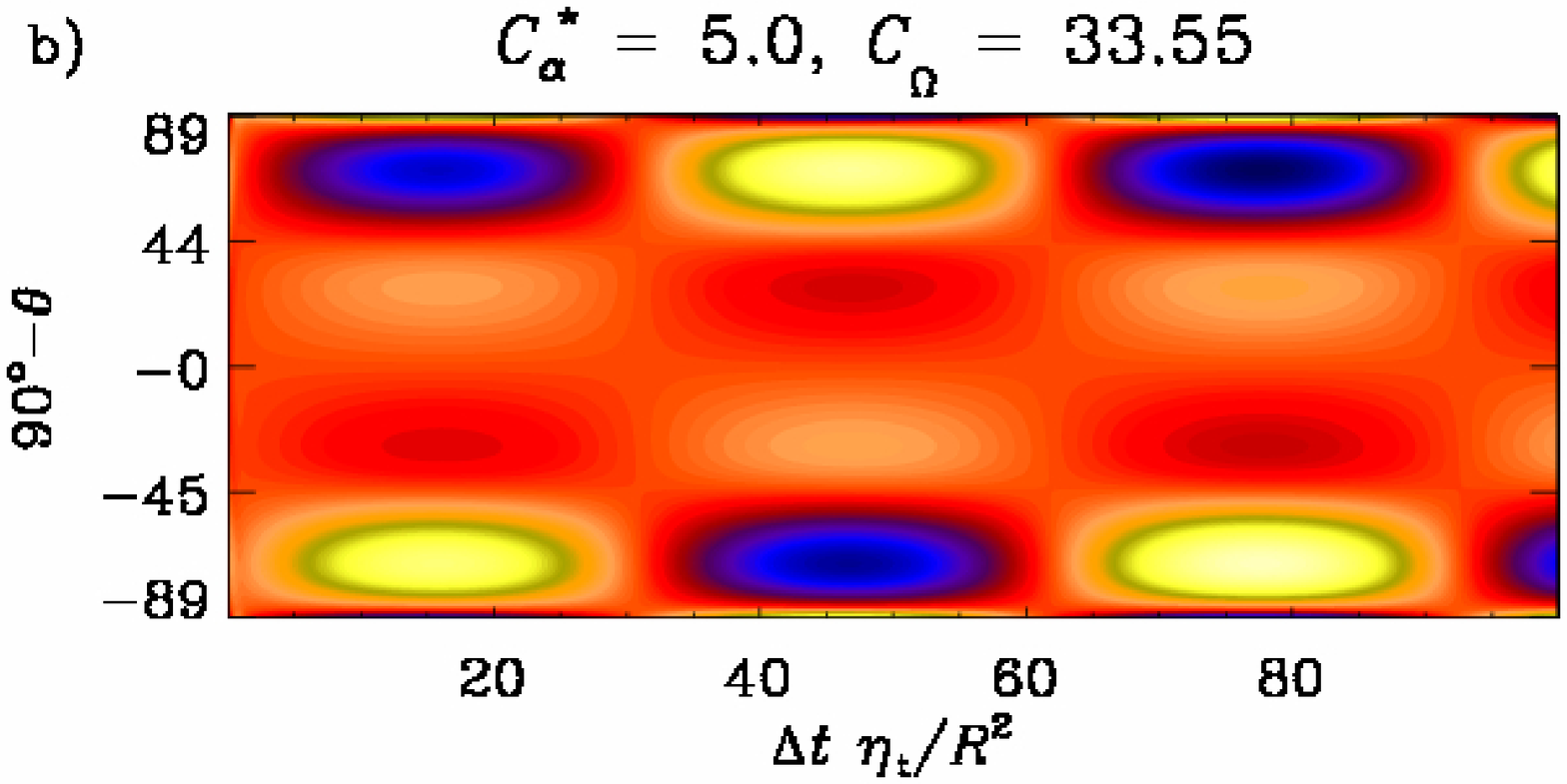}
\includegraphics[width=\columnwidth]{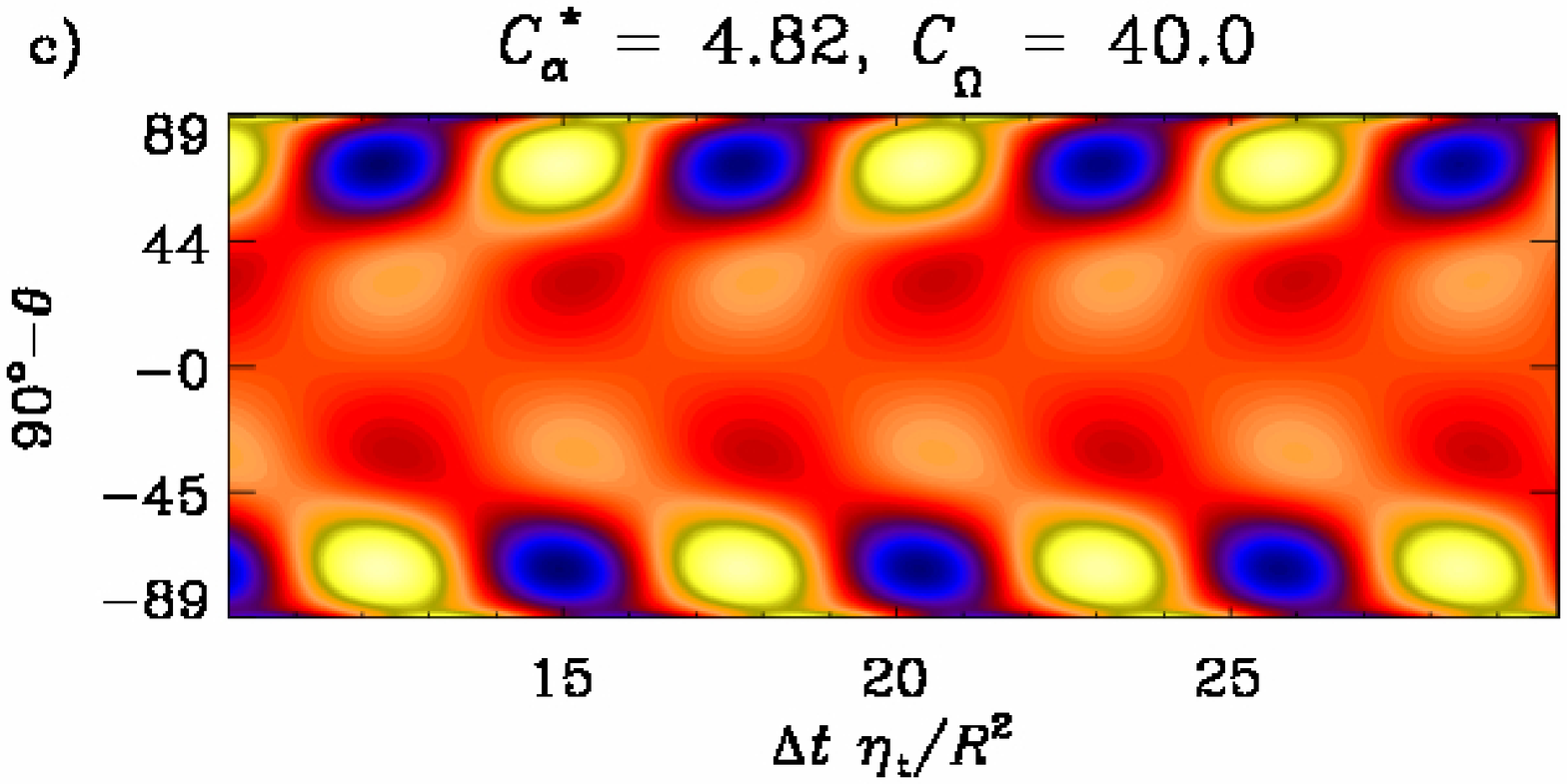}
\includegraphics[width=\columnwidth]{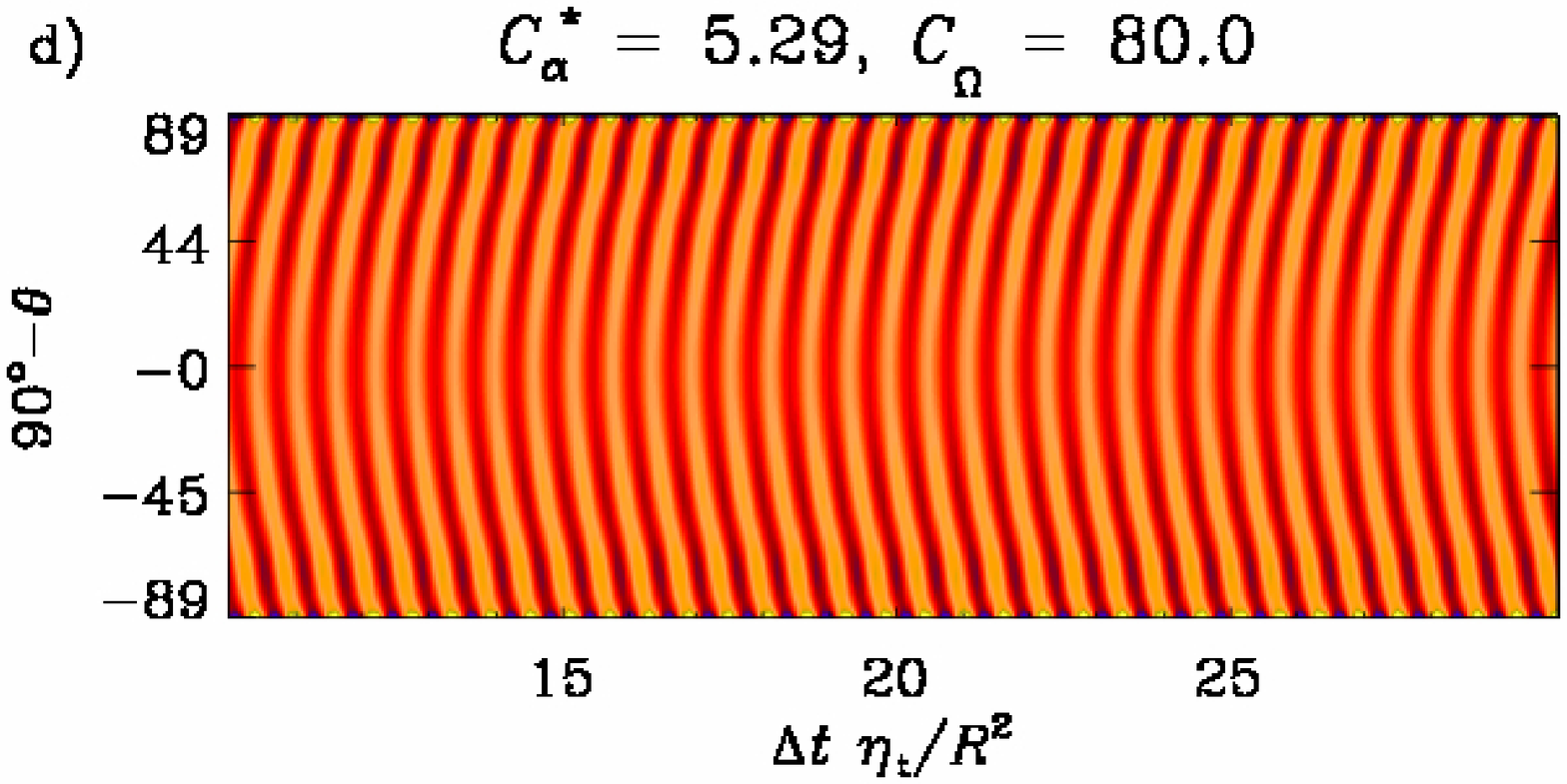}
\includegraphics[width=\columnwidth]{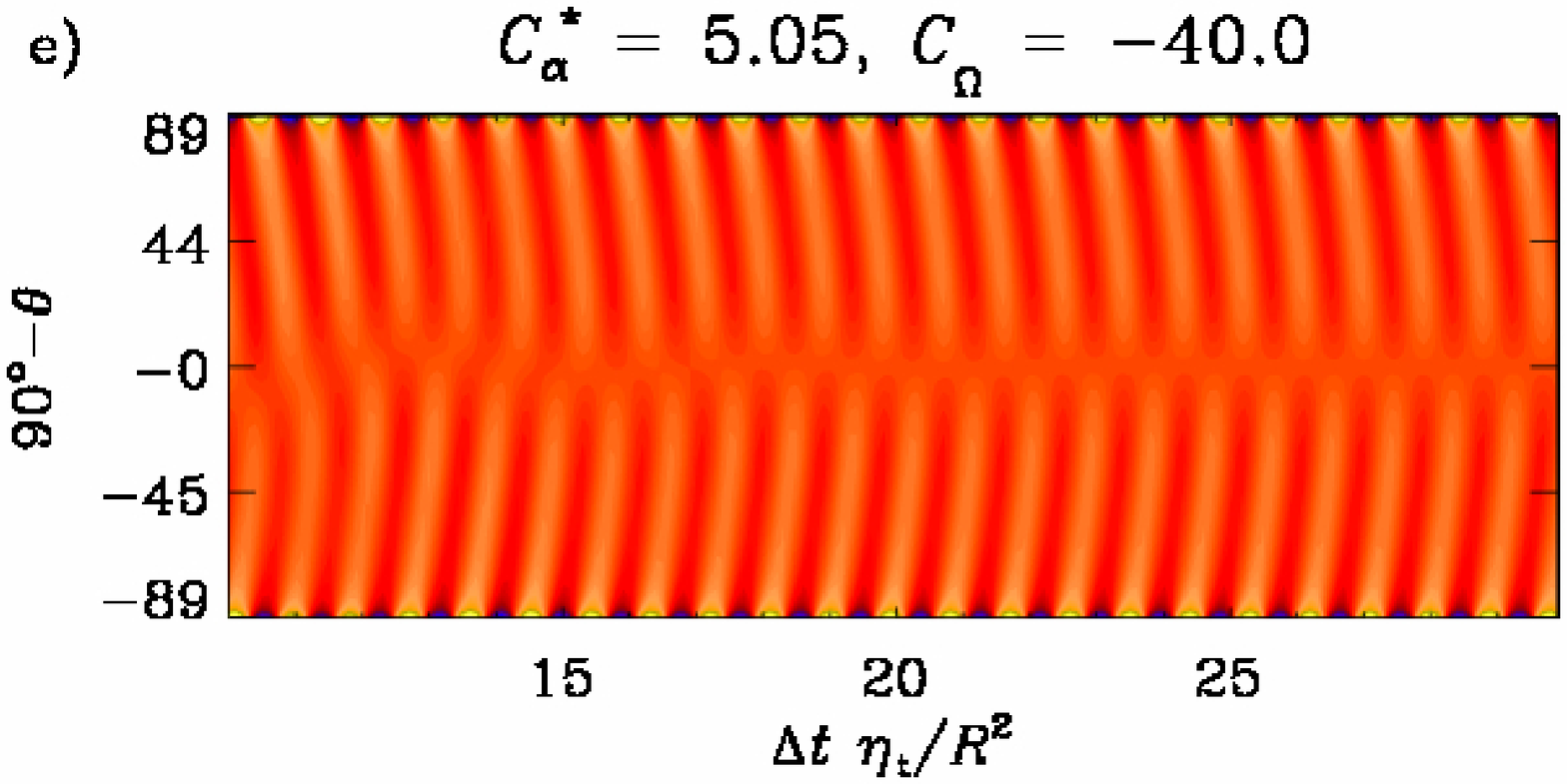}
\end{center}\caption[]{
Azimuthal magnetic field for $\theta_0=1\degr$
with the ASA boundary condition.
}
\label{shear_alpcrit_bfly}
\end{figure}

We calculate the frequency $\omega$ of oscillatory solutions as 
in the previous section and show the results in \Fig{com_freq}. 
It can be seen that for positive shear, 
$\omega$ approaches 0 as $C_\Omega\to 33.55$.
There also exists a jump in frequency around 
$C_\Omega \sim 70$, corresponding to a 
change in the symmetry of the azimuthal field. 
This is demonstrated in \Fig{shear_alpcrit_bfly} where time-latitude
diagrams of the azimuthal magnetic fields are shown for a
representative selection of $C_\Omega$ values for models with
$\theta_0=1\degr$.
The symmetry change corresponding to the 
frequency jump in \Fig{com_freq}
can be seen in the change from antisymmetric 
about the equator (third panel of \Fig{shear_alpcrit_bfly}, $C_\Omega
= 40$) to symmetric (fourth panel of \Fig{shear_alpcrit_bfly},
$C_\Omega = 80$).
The magnetic field is also symmetric in the oscillatory solution found
for $C_\Omega = 0$.

\begin{table}
\caption{
$\Calp$ for runs with $\theta_0=0\degr$, $1\degr$, $5\degr$,
and $15\degr$ with $\aaaa = (1,0,0)$ and $\ee=(1,0,0)$
and $\tilde{\mu}= 1$.
  }
\begin{tabular}{lccc}
\hline \hline
$\theta_0$ & Boundary & $C_\Omega$ & $\Calp$ \\
  & Condition & & \\
  \hline
 0 & SAA & 60 & 5.17 \\
 1 & ASA & 60 & 4.71 \\
 5 & ASA & 60 & 4.18 \\
 15 & ASA & 60 & 4.22 \\ 
 \hline
 \end{tabular}
 \label{shear_vary_theta}
 \end{table}

\begin{figure}[t!]\begin{center}
\includegraphics[width=\columnwidth]{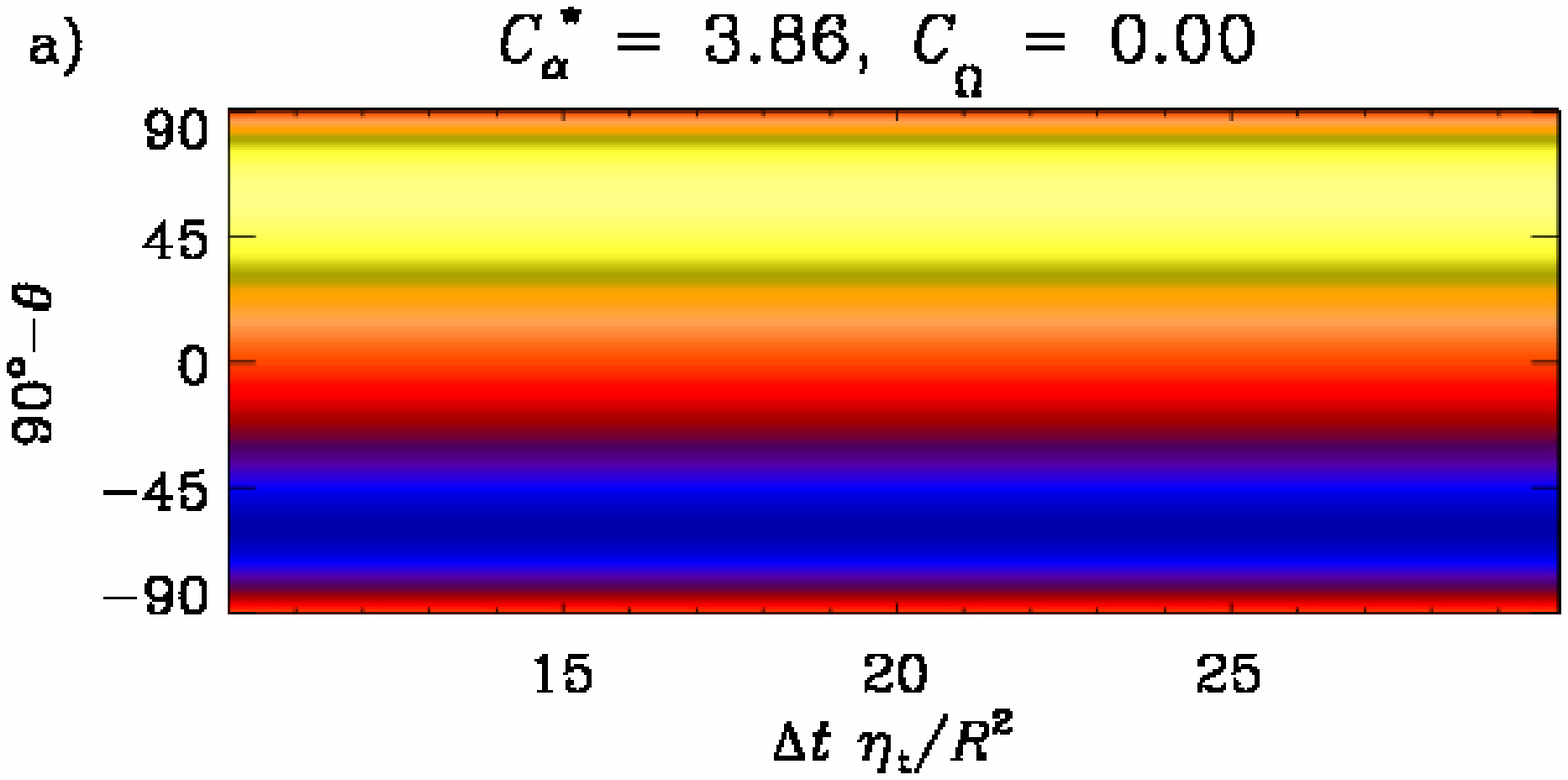}
\includegraphics[width=\columnwidth]{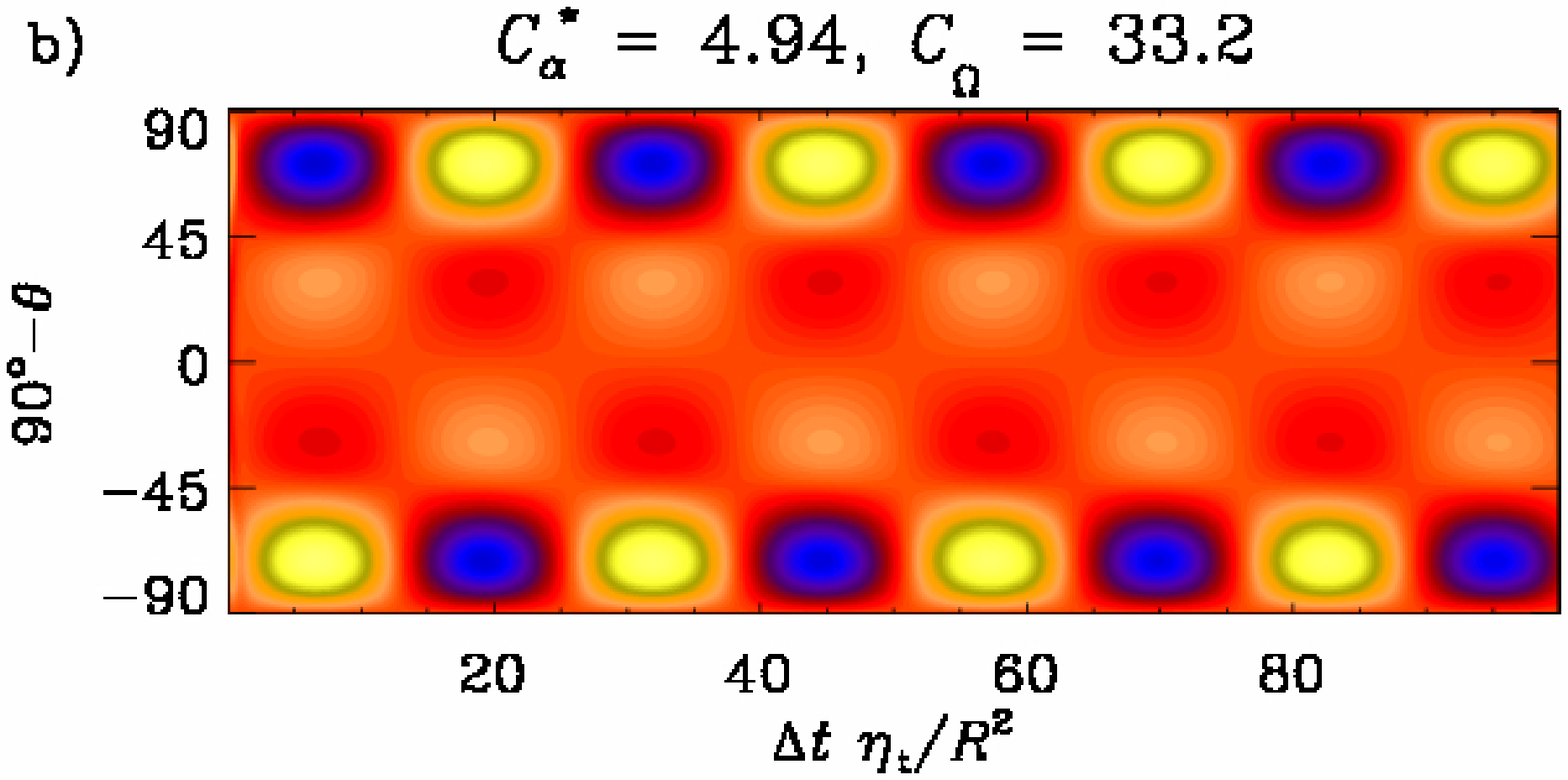}
\includegraphics[width=\columnwidth]{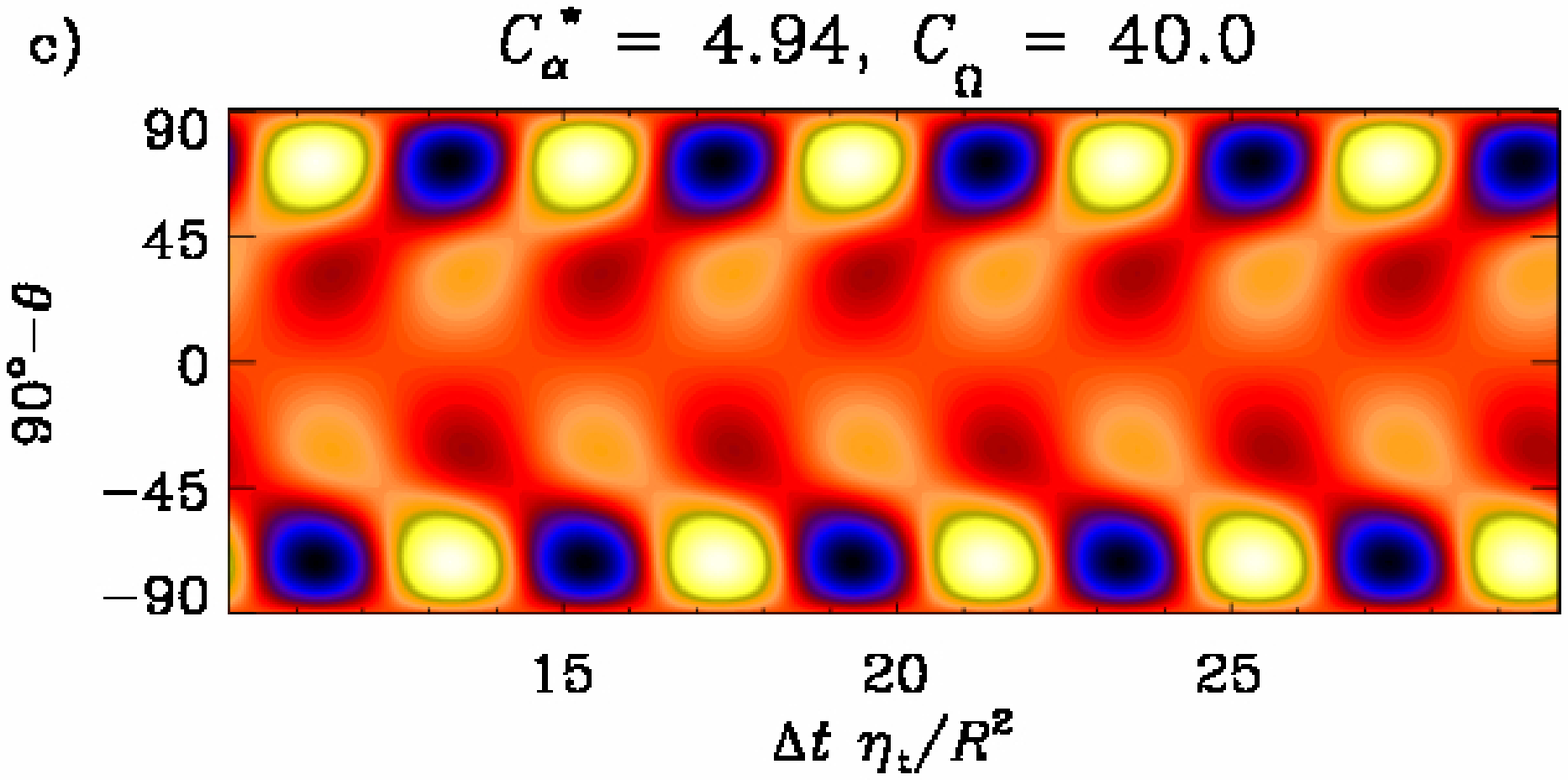}
\includegraphics[width=\columnwidth]{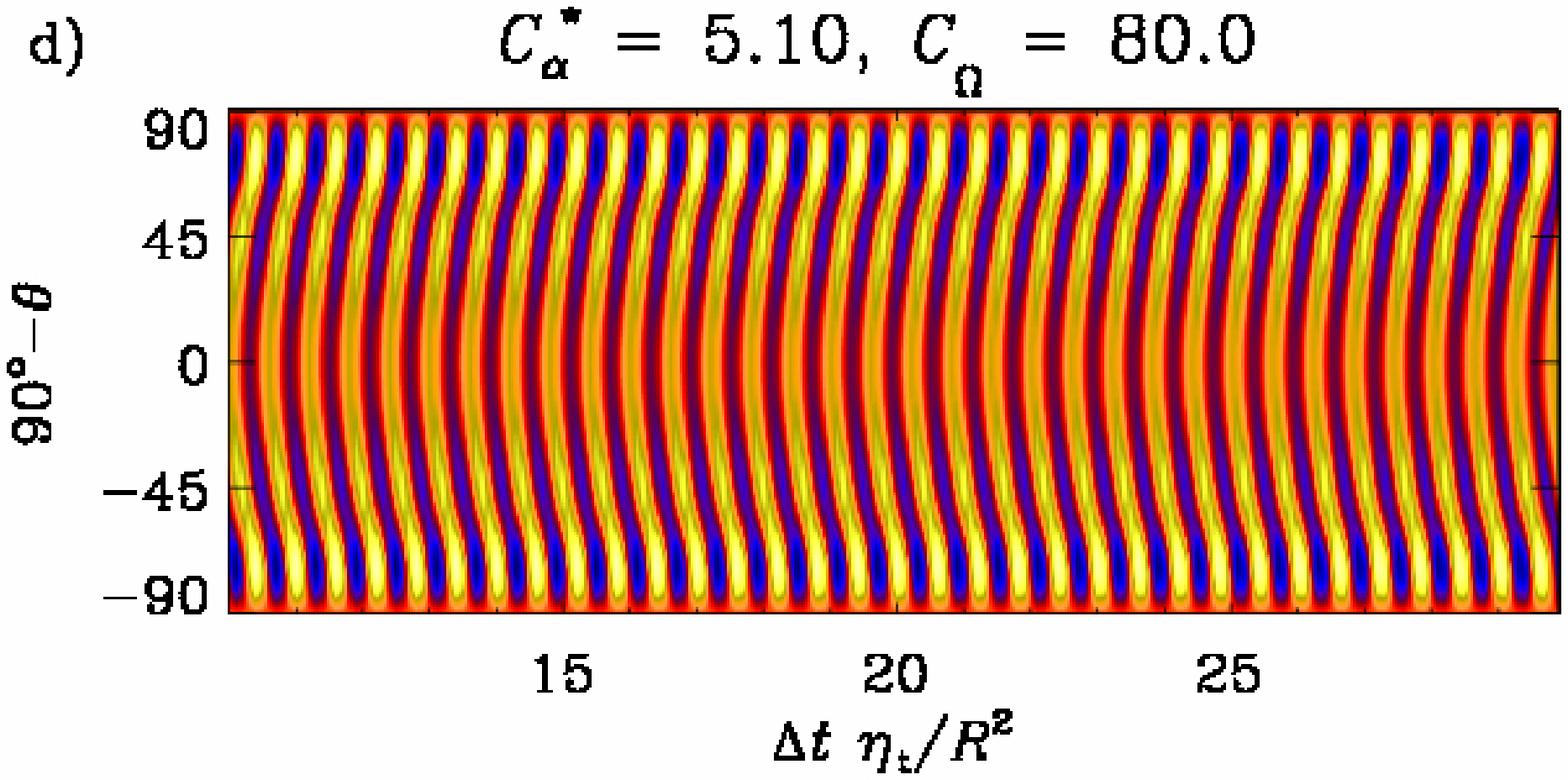}
\includegraphics[width=\columnwidth]{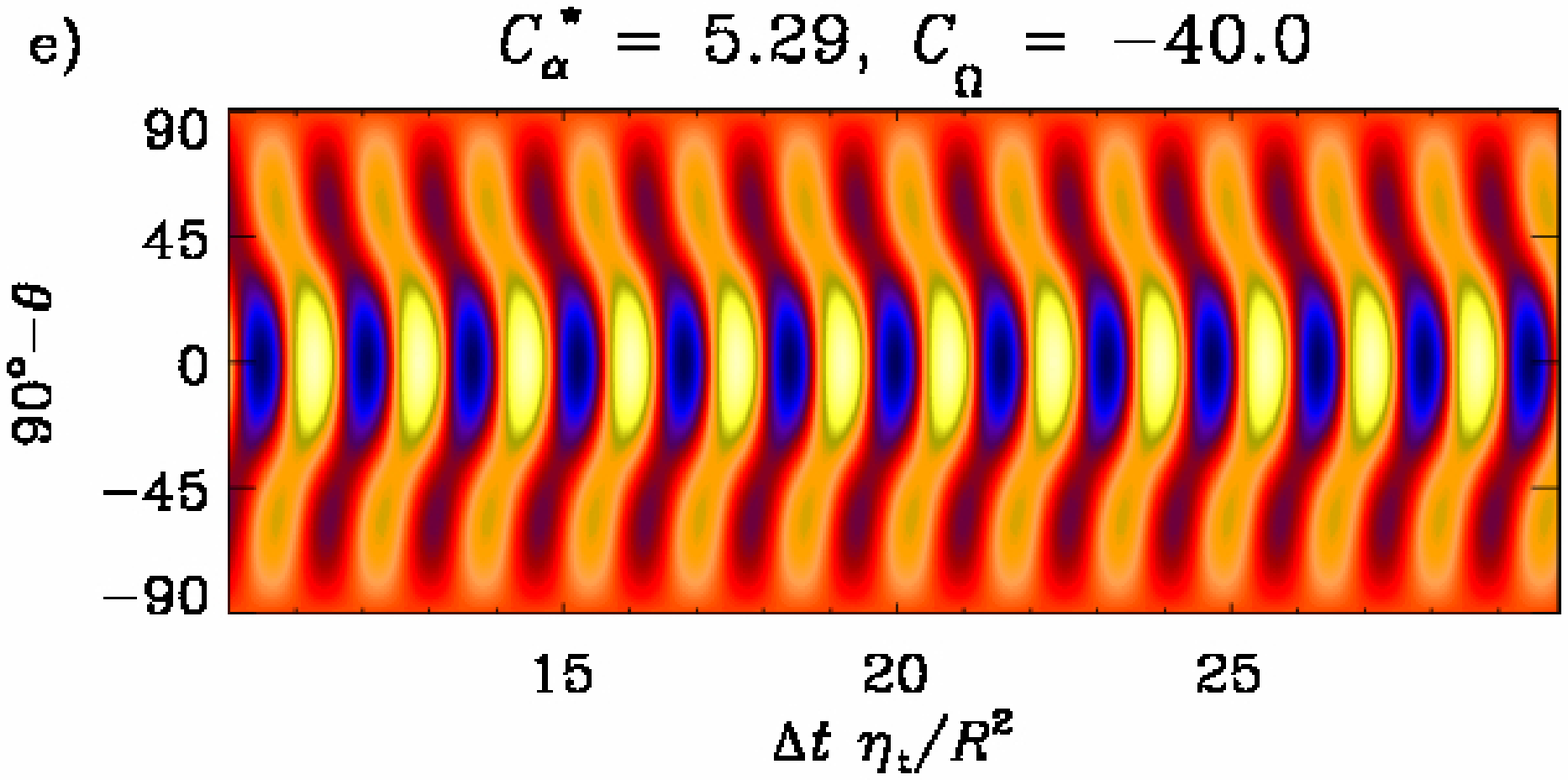}
\end{center}\caption[]{
Azimuthal magnetic field for $\theta_0=0\degr$ 
with the SAA boundary condition.
}
\label{shear_alpcrit_bfly_0deg}
\end{figure}

All oscillatory solutions with positive (negative) shear show poleward
(equatorward) migration in accordance with the Parker--Yoshimura rule
\citep{Par55,Yo75}, see the third and fifth panels of
\Fig{shear_alpcrit_bfly}, respectively, for representative results.
The frequency of the oscillations increases with greater
$C_\Omega$ in accordance with linear theory of $\alpha\Omega$ dynamos,
except that there $|\omega|\propto C_\Omega^{1/2}$ \citep[e.g.][]{BS05}. 
Most of the magnetic field is concentrated at high latitudes above 
$|90\degr-\theta| > 60\degr$ for cases where $C_\Omega$ 
is positive, \Fig{shear_alpcrit_bfly}(b)--(d).
When $C_\Omega \leq 0$, the field is even more concentrated close to
boundaries, \Fig{shear_alpcrit_bfly}(e).

\begin{figure}[t!]\begin{center}
\includegraphics[width=\columnwidth]{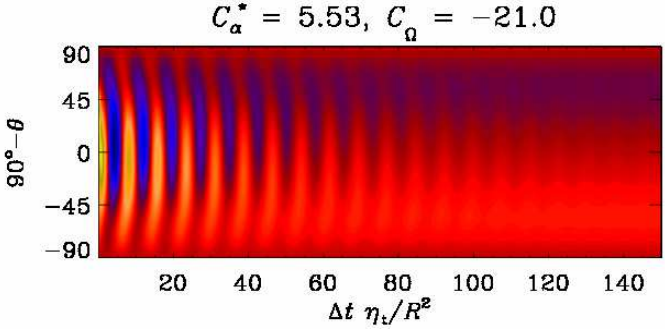}
\end{center}\caption[]{
Azimuthal magnetic field for $\theta_0=0\degr$ 
with the SAA boundary condition.
}\label{shear_bfly_0deg_negcrit}
\end{figure}

\subsubsection{Comparison between $\theta_0=0$ and $\theta_0=1\degr$ cases}

The model is now extended to the poles to study the differences
between wedges and full spheres.
The boundary condition on $\theta_0=0\degr$ 
is changed to comply with the regularity requirement (SAA). We focus
on the case where $\tilde{\mu} = 1$. We
consider a few models with $\tilde{\mu}=0$ and $2$ to probe whether the
behaviour is similar, as in the $\theta_0=1\degr$ case.
We find that the values of $\Calp$ are fairly close to those obtained for
the corresponding $\theta_0=1\degr$ models; see the red symbols in
\Fig{shear_alpcrit}.
Similarly as in the $\theta_0=1\degr$ case, 
a bifurcation into stationary and oscillatory solutions exists in the
positive $C_\Omega$ regime
with a cut-off point at $C_\Omega \approx 33.2$, which is slightly
lower than in the $\theta_0=1\degr$ case.
For negative shear, unlike for $\theta_0=1\degr$
where all values produce oscillatory dynamos, 
the regime for oscillations is found only for $C_\Omega \la -21$.
The oscillatory mode gradually disappears and only a stationary 
mode persists, which is shown in \Fig{shear_bfly_0deg_negcrit}. 

The oscillation frequencies (\Fig{com_freq}, red symbols) are similar
to those in the case of positive shear. Similarly to the
$\theta_0=1\degr$ case, a jump in frequency is observed when the
azimuthal field changes symmetry with respect to the equator, as shown
in \Fig{shear_alpcrit_bfly_0deg}(c) and (d) for
antisymmetric $(C_\Omega = 40)$ and symmetric $(C_\Omega = 80)$ field
configurations, respectively.  In the antisymmetric regime, the
azimuthal field is concentrated at approximately the same latitudes as
for the case $\theta_0=1\degr$.
In the symmetric regime, i.e.\ for $C_\Omega\gtrsim70$, 
the azimuthal field extends to lower latitudes, 
$|90\degr-\theta| > 30\degr$; see 
\Fig{shear_alpcrit_bfly_0deg}(d),
The main difference occurs at the boundary itself such that for
$\theta_0=1\degr$ (ASA) the magnetic field
peaks at the boundary whereas it vanishes at the pole for $\theta_0=0$
(SAA).
When shear is negative, the field instead becomes concentrated 
and symmetric around the equator, 
and in accordance with the Parker--Yoshimura rule,
the dynamo has an equatorward drift. 
The case of negative shear results in a dramatically different 
concentration of the azimuthal field when compared with the
$\theta_0=1\degr$ counterpart; see the bottom panels of
\Figs{shear_alpcrit_bfly}{shear_alpcrit_bfly_0deg} for runs with
$C_\Omega=-40$ for the two cases.
Even though the values for $\Calp$ are similar for 
$\theta_0=0\degr$ and $1\degr$, the frequency of 
oscillations is less by about a factor of two in the former case, see
\Fig{com_freq}.

Finally, we examine the effect that $\theta_0$ has on the results
by holding $C_\Omega$ constant and determining $\Calp$.
The results are given in \Tab{shear_vary_theta}.
We find that there is a dependency on $\theta_0$, but
the behaviour is consistent if one goes to the poles and changes
the boundary condition; see \Tab{shear_vary_theta}
where the change between $\theta_0 = 5\degr$ and 
$1\degr$ is comparable to the difference between 
$1\degr$ and $0\degr$.
All solutions are oscillatory with poleward migration.

Our results suggest that, at least in the 
cases where $C_\Omega > 0$, 
a setup with $\theta_0=1\degr$ and perfect conductor boundary
condition (ASA) gives similar results as full sphere models with
$\theta_0=0\degr$ and the regularity (SAA) condition.
Furthermore, solutions for $33.4 < C_\Omega < 75$ 
are also fairly similar.
This indicates that the wedges are a fair approximation of full
spheres in this parameter regime.
If the shear is negative, there is a qualitative change in the results
between $\theta_0=1\degr$ and $\theta_0=0\degr$ cases.
It appears that oscillatory solutions are obtained only for the ASA
boundaries for weak negative shear.

\subsubsection{Varying the $\alpha$ and $\etat$ profiles}

\begin{figure}[t!]\begin{center}
\includegraphics[width=\columnwidth]{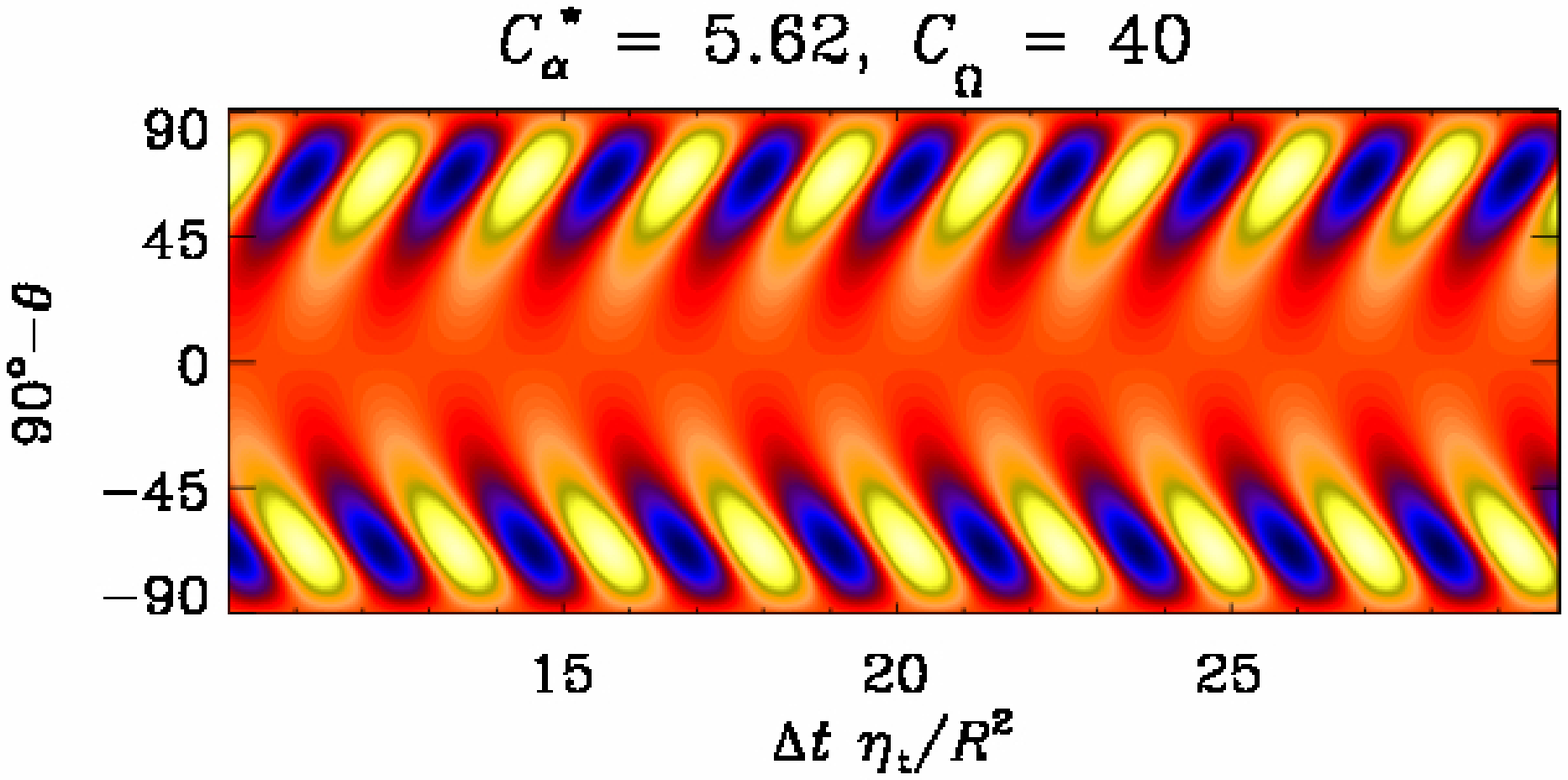}
\includegraphics[width=\columnwidth]{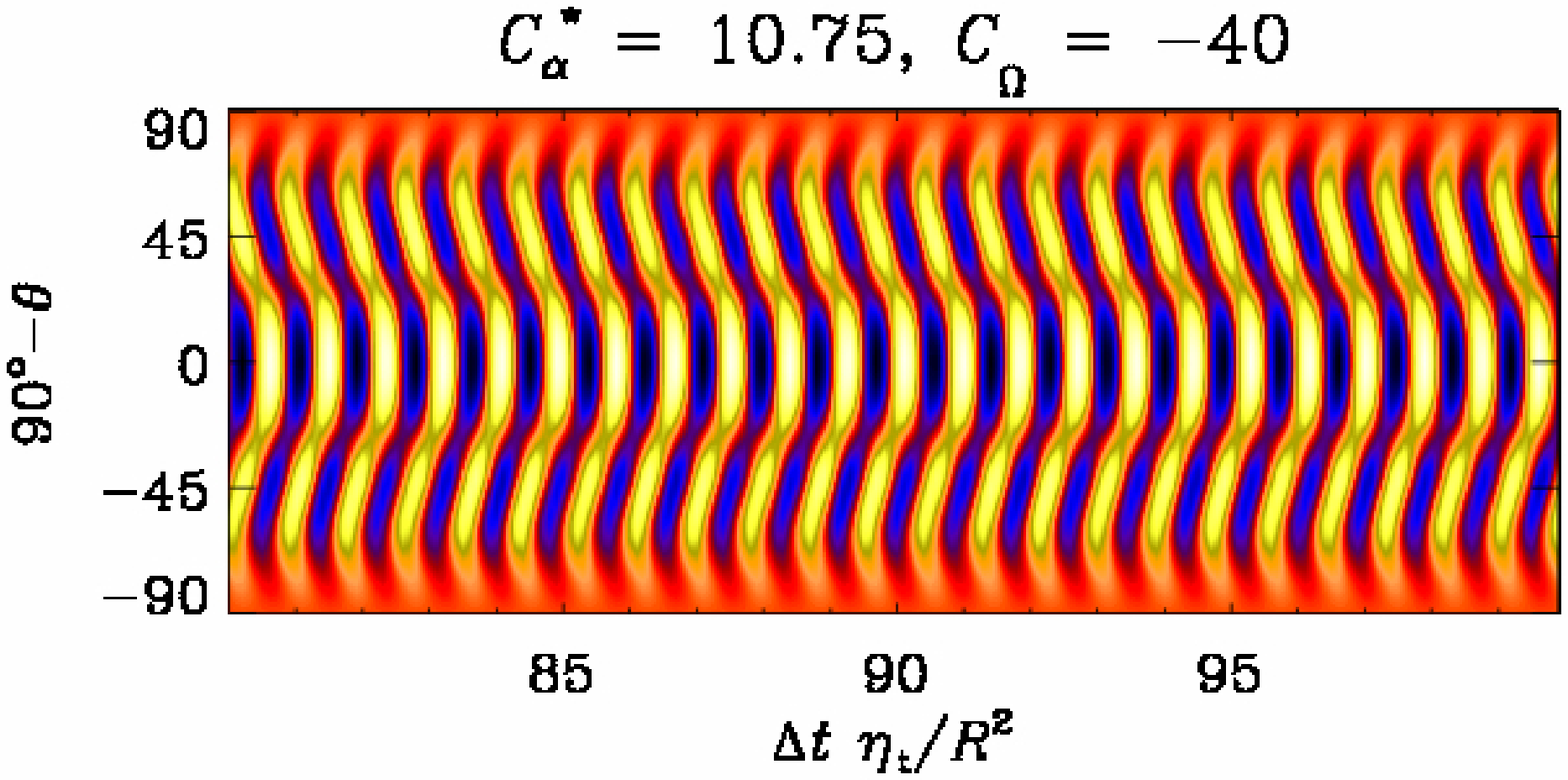}
\end{center}\caption[]{
Azimuthal magnetic field for $\theta_0=0\degr$ 
with $e_0 = 0.05$, $\tilde{\mu} = 1$, and $\aaaa=\ee=(0,0,1)$
for $C_\Omega=40$ (upper panel), and $C_\Omega=-40$ (lower panel).
}
\label{shear_cond}
\end{figure}

Finally, we consider changes to the turbulent magnetic diffusivity profile. 
We do not
perform a thorough parameter study but consider a pair of cases
corresponding to $C_\Omega = \pm40$, $\aaaa=\ee=(0,0,1)$, $e_0
= 0.05$, $\tilde{\mu}=1$, and $\theta_0=0$ with regularity conditions
for the magnetic field. We show the time--latitude diagrams of the
azimuthal field from these models in \Fig{shear_cond}.

In the case of positive shear, the combination of shear, 
$\alpha$ and $\etat$ profiles, creates a steady migration poleward at
latitudes above $\pm45\degr$.
Comparing this to an $\alpha^2$ dynamo with the same profiles of
$\alpha$ and $\etat$ (bottom panel of \Fig{bfly_0deg_sin4y_1}), and to
an $\alpha^2\Omega$ run with no $\sin^{2n}\theta$ contributions in the
profiles but the same value of $C_\Omega$ (third panel of
\Fig{shear_alpcrit_bfly_0deg}), shows that the migration direction is
reversed in comparison to the $\alpha^2$ run and that the poleward
drift is more coherent than in the $\alpha^2\Omega$ model.
These results indicate that the shear 
determines the direction of the dynamo wave in this parameter regime.
The azimuthal field in both of the comparison cases is antisymmetric,
and this result also carries over to the case when shear is included
with the same $\alpha$ and $\etat$ profiles.
The frequency of the oscillations is $\omega=5.54$, 
and the critical dynamo parameter is $\Calp = 5.62$.
These values are somewhat close to the values ($\omega=3.90$ and
$\Calp=4.94$) obtained in Sect.~\ref{sec:overall} in the case with
more uniform profiles of the turbulent transport coefficients.

We found earlier that, in the case of negative shear, the azimuthal
field was symmetric about the equator; see the bottom panel of
\Fig{shear_alpcrit_bfly_0deg}.
With more equatorially concentrated turbulent diffusivity and $\alpha$
profiles we also find solutions with equatorial symmetry, see the
bottom panel of \Fig{shear_cond}.
Furthermore, the magnetic field now has a minimum
around latitudes $\pm25\degr$.  The Parker--Yoshimura rule still holds
true, and the migration is equatorward.
However, $\Calp$ has almost doubled from $5.62$ to $10.75$, and the 
frequency of oscillations is much larger, 
$\omega = 14.56$ in comparison to $5.54$.
The main effect from the more concentrated profiles for $\alpha$ and
$\etat$ in the case of $\alpha^2\Omega$ dynamos is seen in the
latitudinal profile of the resulting magnetic fields, but the
qualitative character of the solutions remains unchanged in comparison
to models with simpler latitude dependence of the turbulent transport
coefficients.

\section{Conclusions}

Motivated by earlier results of global simulations in wedge geometry,
we have studied the robustness of oscillatory solutions in
$\alpha^2$ dynamos in simple one-dimensional mean-field dynamo models.
We found that the boundary conditions on the latitudinal boundaries play
a major role in the realised solutions for $\alpha^2$ dynamos with a
simple $\cos\theta$ profile for $\alpha$ and constant turbulent
diffusivity.
Imposing the 
perfect conductor boundary condition creates oscillating solutions only 
for solutions where $\theta_0 \ga 1\degr$. 
For $\theta_0=1\degr$, both oscillatory and 
stationary solutions were found to appear with slightly differing
critical dynamo numbers.
We found no oscillatory solutions for the normal field (SAS) or
regularity conditions (SAA).

Keeping a simple $\cos\theta$ profile for the $\alpha$ effect and
varying the $\etat$ profile creates oscillating solutions with a low
frequency and no clear migration or stationary solutions, depending on
the value of the underlying (constant) magnetic diffusivity.
The magnetic field is largely concentrated near the poles. 
If the $\alpha$ profile is changed to be concentrated
near the equator, similar to profiles observed 
in rapidly rotating turbulent convection, 
the magnetic field
becomes more evenly distributed towards the equator.
The magnetic field also exhibits clear equatorward migration and
antisymmetry with respect to the equator.
The overall conclusion is that $\alpha^2$ dynamos can produce
solar-like magnetic activity if the $\alpha$ effect and turbulent
diffusivity have latitudinal profiles that are sufficiently concentrated toward
the equator.

We then added positive shear to study $\alpha^2\Omega$ dynamos and how
they connect to the pure $\alpha^2$ solutions in the same wedge geometry
with $\theta_0=1\degr$.
For weak shear the azimuthal magnetic field is 
concentrated at the poles and shifts equatorward.
Over a certain interval in $C_\Omega$, which depends on the added local
friction $\tilde{\mu}$, oscillatory solutions are found and
the field is more concentrated 
across all upper latitudes.
For $\tilde{\mu}=1$, we found that both stationary and oscillatory
solutions exist with the oscillatory one having a substantially
higher critical dynamo number.
Going to a full sphere with $\theta_0 = 0\degr$ and changing the
boundary condition to SAA produced qualitatively and quantitatively
similar results when the shear was positive.
Results are less similar if negative shear is introduced.
When $\theta_0=1\degr$, all solutions found where 
$C_\Omega < 0$ were found to oscillate. 
However, when $\theta_0 = 0\degr$, shear had to 
exceed a critical value,
$|C_\Omega| = 21$,
for solutions to oscillate. 
Furthermore, the structure of the azimuthal field 
over time was significantly different, showing 
symmetry about the equator and concentration at the equator. 
In all cases with shear, the Parker--Yoshimura 
rule was found to be obeyed where oscillatory solutions with 
negative shear migrated equatorward and 
positive shear, poleward.

When combining the $\etat$ profile 
with shear, the direction of migration was determined 
by the sign of $C_\Omega$.
The frequency increased in the case of negative
shear when using an $\etat$ and $\alpha$ profile 
with higher order terms.   

There are other possibilities for refining the model and for
obtaining oscillatory solutions to the $\alpha^2$ dynamo.
One possibility is to study the effect of decreasing
the (microphysical) magnetic diffusivity even further.
Another possibility is to study the memory effect, which has recently been identified
as a means to facilitate oscillatory behaviour, although so far only decaying
solutions have been found to be modified in that way \citep{DBM13}.
However, under suitable conditions such solutions can indeed become
oscillatory \citep{RDRB14} and may present a possible solution 
to the problem where equatorward motion obtained via 
varying the $\alpha$ profile (i.e., $\aaaa = (0,0,1)$) 
is limited to certain latitudes.


\begin{acknowledgements}
  The authors thank Nordita for hospitality during their visits.
  Financial support from the Vilho, Yrj\"o and Kalle V\"ais\"al\"a
  Foundation (EC), the Academy of Finland grants No.\ 136189, 140970
  (PJK) and the Academy of Finland Centre of Excellence ReSoLVE
  (272157; MJK and PJK), as well as the Swedish Research Council grants
  621-2011-5076 and 2012-5797, and the European Research Council under
  the AstroDyn Research Project 227952 are acknowledged.  We
  acknowledge CSC -- IT Center for Science Ltd., who are administered
  by the Finnish Ministry of Education, for the allocation of
  computational resources.

\end{acknowledgements}


\vfill\bigskip\noindent\tiny\begin{verbatim}

\end{verbatim}
\end{document}

%% file: macros.tex
\newcommand{\EQ}{\begin{equation}}
\newcommand{\EN}{\end{equation}}
\newcommand{\EQA}{\begin{eqnarray}}
\newcommand{\ENA}{\end{eqnarray}}

\newcommand{\Eq}[1]{Eq.~(\ref{#1})}
\newcommand{\Eqs}[2]{Eqs.~(\ref{#1}) and~(\ref{#2})}

\newcommand{\Fig}[1]{Fig.~\ref{#1}}

\newcommand{\Figs}[2]{Figs.~\ref{#1} and \ref{#2}}

\newcommand{\Tab}[1]{Table~\ref{#1}}

\newcommand{\mean}[1]{\overline #1}

{}
{}
{}

{}
{}
\newcommand{\meanEMF}{\overline{\mbox{\boldmath ${\cal E}$}}{}}{}
{}
{}
{}
{}
{}
\newcommand{\meanAA}{\overline{\mbox{\boldmath $A$}}{}}{}
\newcommand{\meanBB}{\overline{\mbox{\boldmath $B$}}{}}{}
{}
{}
{}
{}
{}
{}
{}
{}
{}
\newcommand{\meanJJ}{\overline{\mbox{\boldmath $J$}}{}}{}
{}
\newcommand{\meanUU}{\overline{\mbox{\boldmath $U$}}{}}{}

{}
{}
{}
\newcommand{\meanA}{\overline{A}}

\newcommand{\meanU}{\overline{U}}

{}

{}
{}
{}

\newcommand{\pphi}{\hat{\bm{\phi}}}

\newcommand{\aaaa}{\mbox{\boldmath $a$} {}}
\newcommand{\ee}{\mbox{\boldmath $e$} {}}

\newcommand{\grav}{\mbox{\boldmath $g$} {}}
\newcommand{\nab}{\mbox{\boldmath $\nabla$} {}}

\newcommand{\OO}{\bm{\Omega}}

\def\degr{\hbox{$^\circ$}}
\def\la{\mathrel{\mathchoice {\vcenter{\offinterlineskip\halign{\hfil
$\displaystyle##$\hfil\cr<\cr\sim\cr}}}
{\vcenter{\offinterlineskip\halign{\hfil$\textstyle##$\hfil\cr<\cr\sim\cr}}}
{\vcenter{\offinterlineskip\halign{\hfil$\scriptstyle##$\hfil\cr<\cr\sim\cr}}}
{\vcenter{\offinterlineskip\halign{\hfil$\scriptscriptstyle##$\hfil\cr<\cr\sim\cr}}}}}
\def\ga{\mathrel{\mathchoice {\vcenter{\offinterlineskip\halign{\hfil
$\displaystyle##$\hfil\cr>\cr\sim\cr}}}
{\vcenter{\offinterlineskip\halign{\hfil$\textstyle##$\hfil\cr>\cr\sim\cr}}}
{\vcenter{\offinterlineskip\halign{\hfil$\scriptstyle##$\hfil\cr>\cr\sim\cr}}}
{\vcenter{\offinterlineskip\halign{\hfil$\scriptscriptstyle##$\hfil\cr>\cr\sim\cr}}}}}
\def\Calp{C_\alpha ^\star}

\def\etat{\eta_{\rm t}}
\def\etatz{\eta_{\rm t0}}
\def\etatz{\eta_{\rm t0}}

\newcommand{\yapj}[3]{ #1, {ApJ,} {#2}, #3}

\newcommand{\yapjl}[3]{ #1, {ApJ,} {#2}, #3}

\newcommand{\yan}[3]{ #1, {Astron.\ Nachr.,} {#2}, #3}

\newcommand{\yana}[3]{ #1, {A\&A,} {#2}, #3}

\newcommand{\ygafd}[3]{ #1, {Geophys.\ Astrophys.\ Fluid Dyn.,} {#2}, #3}

\newcommand{\yjfm}[3]{ #1, {J.\ Fluid Mech.,} {#2}, #3}

\newcommand{\ypp}[3]{ #1, {Phys.\ Plasmas,} {#2}, #3}

\newcommand{\yprt}[3]{ #1, {Phys.\ Rep.,} {#2}, #3}

\newcommand{\ymn}[3]{ #1, {MNRAS,} {#2}, #3}

\newcommand{\yjour}[4]{ #1, {#2}, {#3}, #4}

